\documentstyle[aps,epsfig]{revtex}
\begin{document}
\title{Non-Markovian quantum trajectories for spectral detection}  
\author{M. W. Jack\footnote{email:mwj@phy.auckland.ac.nz}, M. J. Collett and D. F. Walls}
\address{Department of Physics, University of Auckland, Private Bag 
92019, Auckland, New 
Zealand}
\maketitle 
\begin{center}
(\today)
\end{center}
\begin{abstract}
We present a formulation of non-Markovian quantum trajectories 
for open systems from a measurement theory perspective.
In our treatment there are three distinct ways in which non-Markovian behavior can 
arise; a mode dependent coupling between bath (reservoir) and system,
 a dispersive bath, and by spectral 
detection 
of the output into the bath. In the first two cases the non-Markovian 
behavior is intrinsic to the interaction, in the 
third case the non-Markovian behavior arises from the method of 
detection.  We focus in detail on the trajectories which simulate real-time 
spectral detection of the light emitted from a localized system. 
In this case, the non-Markovian behavior 
arises from the uncertainty in the time of emission 
of particles that are later detected. The results of computer 
simulations of the spectral detection of the 
spontaneous emission from a strongly driven 
two-level atom are presented. 
\end{abstract}
\vspace{1cm}				 
																					  
Quantum trajectories or stochastic Schr\"{o}dinger 
equations \cite{dalibar,carmichael,gisin} have been used as an 
effective numerical tool  (for example \cite{gardiner,molmer,dum,tian}) to 
solve the time evolution of the reduced density matrix for an open 
system. In addition, a single quantum trajectory simulates the evolution of a 
system undergoing continuous measurement of its environment 
\cite{carmichael,photon,wiseman,goetsch}. This is of fundamental 
importance in understanding open system behavior as features of a single 
realization of a measurement process can be obscured 
by density matrix methods which average over the individual realizations
(see \cite{javanainen} for a dramatic example).

Traditionally, quantum trajectories have described 
Markov processes, where the conditioned state of the system at a 
certain time contains all the information needed to calculate the measurement 
probabilities over the next infinitesimal interval. In this case, 
a master equation can describe 
the evolution of the reduced density matrix for the open system. 
Recently, however, there has been  interest  in studying open 
systems for which the Markov approximation  cannot be made. For 
example, in a 
photonic band-gap material the 
correlation time of the electromagnetic vacuum fluctuations near a band edge 
is not small on the timescale of the 
emitting atomic system \cite{bay,vats}.  As another example, a Bose-Einstein 
condensate  losing atoms due to a radio-frequency output 
coupler \cite{outputcoupler} 
experiences Non-Markovian decay for certain coupling rates, 
as the dispersion relation for 
atoms in free space leads to a non-vanishing correlation time for 
the atoms coupled out of the
condensate \cite{hope,savage}.  
There are also measurement processes that lead to a non-Markovian 
evolution of the conditioned system \cite{caves,barchielli}.
Real-time spectral measurement is an example \cite{wisemanthesis}. 
In this case, the measurement process yields not 
only temporal information but also frequency information 
about the system \cite{cohen,cresser}. 
It is therefore of interest to generalize quantum trajectory 
techniques to deal with these non-Markovian processes 
\cite{imamoglu,diosi97,diosi98}. In the non-Markovian case the 
future measurement probabilities need to be calculated from the 
evolution of the  system in 
the past.
 
In this paper we formally derive in Sec.\ref{one} the evolution equation 
 for a non-Markovian 
trajectory for arbitrary measurement schemes. As an illustrative example 
we investigate in Sec.\ref{two}
 the 
trajectories that arise from spectral detection of the light from a 
strongly driven two-level atom.

 A derivation of non-Markovian 
quantum trajectories from a microscopic model (in the case of heterodyne 
detection \cite{wisemanhetrodyne}) has been made recently by Di\`{o}si and Strutz 
\cite{diosi97} via path integrals. By rewriting the path integral 
evolution in terms 
of operators Di\`{o}si {\em et al}. \cite{diosi98} have used 
this formulation to simulate trajectories for some simple 
systems with the requirement that the trajectories are consistent with the 
general formulation.  It is not clear (at least to the present authors) what 
physical situations these trajectories correspond to.
In contrast, our formulation is from a measurement theory perspective. 
We consider a microscopic model and general measurement schemes 
 that are in principle physically realizable. The trajectories are 
defined from this microscopic model and have an obvious interpretation 
in terms of the measurements.

A  non-Markovian {\em quantum} process is one in which the measurement 
probabilities in the present depend
 on the evolution of the system 
in the past. 
This arises naturally when considering the evolution of an open 
system 
conditioned on measurements made on its emission at a distant 
detector. 
In a photon counting experiment, for example, the 
probability of detecting a photon at the detector can, in general, be expressed 
as the probability for the system to undergo emissions distributed 
over times in the past.  In the Markov case, the probability of a 
detection is determined by that of a single time of emission.
In turn, a measurement result 
at a particular time does not completely determine the state of the 
system at the same time. In a Markov process our knowledge of the 
state of the system straight after the measurement is complete. In 
the non-Markov case there is a definite evolution that the system 
must have undergone in the past in order to produce the measurement 
result, but there is always the possibility that future measurement 
results will require us to revise our knowledge of the evolution of 
the system over 
this same time. However, a conditioned system state can be defined  a 
finite time in the past if we can assume that any events occurring in 
the system earlier than this time will have a negligible effect on the 
measurement results in the present.

It is of interest to be able to simulate measurement processes that yield not 
only temporal information but also frequency information 
about a system. In other words, real-time 
spectral measurements, 
where the emission from the system is passed through a `spectrometer' 
that splits it into separate spectral components before each 
component 
is measured. Real-time spectral measurement is relevant to 
situations where different spectral components of the emission from a 
system are time correlated. A well-known example is the three peaked spectrum 
of 
the spontaneous emission from a strongly driven two-level atom 
\cite{mollow}. In 
this system the two side peaks are correlated via photon 
 bunching and 
are individually anti-bunched \cite{cohen,aspect}. The temporal features 
are not accessible 
by methods that time average the detections to determine the spectrum.
Here we  will treat this problem from a real-time spectral detection 
perspective. 

There are two complementary approaches to modeling spectral 
detection. The first and 
most physical approach is to model the spectrometer as a separate 
physical system that is being driven by the output from the system of 
interest. The whole extended system of system plus spectrometer can then 
be evolved forward 
together as a cascaded system 
\cite{gardinercascade,carmichaelcascade}. Imamo\={g}lu 
\cite{imamoglu} has 
 recently developed methods using the extended system approach for 
the more general case of a mode-dependent coupling to the bath. The 
second approach, which perhaps 
coincides more closely with a signal processing treatment, is to 
consider the 
spectrometer as a black box where the output from the box is related 
to the input by a spectral 
response.  Here we will consider the second approach as it is more 
general and is, we believe, closer to the idea behind quantum 
trajectories; of 
simulating a continuous measurement process at a distant detector by 
evolving a local system forward in time. However, this approach means 
that we must consider non-Markovian evolution of the system, since 
emission times become more uncertain as the frequency information 
becomes 
more precise.

\section{Non-Markovian Quantum Trajectories}\label{one}
In this first section we outline the general theory of non-Markovian 
quantum 
trajectories from a measurement theory perspective. The form of the 
system-bath coupling that we consider is given in 
Sec.\ref{interaction} and we introduce a complete set of measurement 
channels in Sec.\ref{channels} to model measurements that are sensitive to only a range of 
modes of the bath. In Sec.\ref{memorytime} we discuss the  finite memory-time 
assumption for the system.

\subsection{Measurements on an open system}
Consider the typical open system situation of a small localized 
system surrounded by an infinite bath. 
The Hamiltonian of the bath and system can be written as the 
sum of three parts
\begin{equation}
	H=H_{\rm sys}+H_{\rm bath}+H_{\rm int},
	\label{total hamiltonian}
\end{equation}
where $H_{\rm sys}$ is the Hamiltonian of the free system, $H_{\rm bath}$ the 
Hamiltonian of the free bath and $H_{\rm int}$ the interaction between the 
two.
 
An open system can be thought of in terms of inputs and outputs 
\cite{collett}: the 
initial bath propagates in towards the system; they 
interact; and the bath propagates away again. 
We would like to model the situation where measurements are 
 being made continuously on the output bath. 
The probability of getting a particular string of measurement results
 ${R}(t,t_{0})$  during the 
interval $[t,t_{0})$, denoted here by $P[{R}(t,t_{0})]$, is given in terms of the initial state by
\begin{eqnarray}
	P[R(t,t_{0})] & = & {\rm Tr}\left\{|R(t,t_{0})\rangle\langle 
	R(t,t_{0})|\rho(t)\right\},\\
	&=& {\rm Tr}\left\{|R(t,t_{0})\rangle\langle 
	R(t,t_{0})|U_{\rm int}(t,t_{0})\rho(t_{0})U_{\rm 
	int}^{\dagger}(t,t_{0})\right\},\label{trace}
\end{eqnarray}
where $U_{\rm int}(t,t_{0})$ is the unitary evolution operator (in the 
interaction picture) determined by the 
total Hamiltonian, Eq.(\ref{total hamiltonian}). Note that we will 
often use the 
labels $R$ and $I$ to label both a general element of the set of all 
possible values and also to represent a particular element of the set.
We assume that the system and bath are initially 
uncorrelated and we can write the initial density matrix in the form 
$\rho(t_{0})=\rho_{\rm bath}(t_{0})\rho_{\rm sys}(t_{0})$.  The density matrix 
of the bath can be written
$\rho_{\rm bath}(t_{0})=\sum_{I} P[I(t,t_{0})]|I(t,t_{0})\rangle\langle 
I(t,t_{0})|$ where $I(t,t_{0})$ labels a particular pure state of the 
input bath. Similarly, the system density matrix can be written, $\rho_{\rm sys}(t_{0})=\sum_{k} P[\psi_{k}(t_{0})]|\psi_{k}(t_{0})\rangle\langle 
\psi_{k}(t_{0})|$. We can now write the probability of the result 
$R(t,t_{0})$ in the form
\begin{eqnarray}
	P[R(t,t_{0})] & = & \sum_{I k}P[R(t,t_{0})I(t,t_{0})\psi_{k}(t_{0})],\\
	& =& \sum_{I k} 
	P[R(t,t_{0})|I(t,t_{0})\psi_{k}(t_{0})]P[I(t,t_{0})]
	P[\psi_{k}(t_{0})],\label{conditional}
\end{eqnarray}
where we have used Bayes theorem. From Eq.(\ref{trace}) the conditional probability can 
be written 
\begin{equation}
 	P[R(t,t_{0})|I(t,t_{0})]  = \langle \psi(t_{0})|\Omega^{\dagger}_{R I}(t,t_{0})
 	\Omega_{R I}(t,t_{0})|\psi(t_{0})\rangle,
 \end{equation}
where we have defined the {\em operation} \cite{davies} on the system space corresponding 
to these 
measurement results over the interval $[t,t_{0})$ as 
\begin{equation}
 \Omega_{R I}(t,t_{0})=\langle  
 	R(t,t_{0})|U_{\rm int}(t,t_{0})|{I}(t,t_{0})\rangle.
\end{equation}
Note that we have suppressed the explicit reference to the probability 
distribution over the initial system states for ease of 
notation. This is equivalent to assuming that the system was initially 
in a pure state. 
The explicit form of the operation $\Omega_{RI}$
for a linear, particle-conserving coupling will be given in 
Sec.\ref{interaction}. 
For now, we simply assume that an explicit form exists. 
The idea behind a quantum trajectory simulation is expressed by 
Eq.(\ref{conditional}). If in a number of runs of the simulation we choose the initial
 states randomly with 
a frequency corresponding to the initial 
probability distribution $P[I(t,t_{0})]$  
then the conditional probabilities $	P[R(t,t_{0})|I(t,t_{0})] $ of each run will automatically 
generate  values of $R(t,t_{0})$ consistent with $P[R(t,t_{0})]$. 

In general, the above operation will not satisfy the semi-group property 
characteristic of a Markov process, i.e, 
$\Omega(t,t_{0})\neq\Omega(t,s)\Omega(s,t_{0})$ for all $s$ (we have 
dropped the subscripts for clarity).
However, the operation can always be written in the form 
\begin{equation}
	\Omega(t,t_{0})=\Omega(t,s)\circ\Omega(s,t_{0}),
\end{equation}
where we have defined the $\circ$-product by
\begin{equation}
	\Omega(t,s)\circ\Omega(s,t_{0})=\sum_{k}\Omega^{k}(t,s)\Omega^{k}(s,t_{0}).
\label{dotproduct}
\end{equation}
This is because one can always insert a sum over a complete set of 
states, $\openone=\sum |k\rangle\langle k|$, at any time in 
the unitary evolution. The summation in 
Eq.(\ref{dotproduct}) 
need not be over the complete set of states, but is over a subset of 
these that are consistent with both the input and resultant 
states and therefore depends on the time $s$. The inability to 
factorize  this operation is due to the fact that the system is
entangled with the bath at time $s$, so that a 
measurement of the output at this time does not completely determine the state of 
the system at the same time. 

To understand this we need to consider in more detail the differences between
Markovian and non-Markovian evolution of open systems.
Firstly,  it is not 
possible to write down a master equation for the reduced system density matrix
$\rho_{\rm sys}(t)={\rm Tr}_{\rm bath}\{\rho(t)\}$. 
This is evident even to first order perturbation in the interaction 
Hamiltonian during the time interval
$T_{\rm m}$,
\begin{equation}
	\rho(t)-\rho(t-T_{\rm m})=\int^{t}_{t-T_{\rm m}}ds[H_{\rm 
	int}(s),\rho(t-T_{\rm m})].
	\label{density matrix}
\end{equation}
where $H_{\rm int}(s)$ is in the interaction picture and we have 
assumed that the contributions from the state at times before $t-T_{\rm 
m}$ are negligible.
We now write the density matrix at time $t-T_{\rm m}$ in terms of a 
factorized part and an unfactorized part $\rho(t-T_{\rm 
m})=\rho^{0}_{\rm sys}(t-T_{\rm m})\rho_{\rm bath}(t_{0})+\chi(t-T_{\rm 
m})$, where $\chi(t-T_{\rm m})$ contains the entanglement. Inserting this into the above equation and taking the trace over 
the bath variables we find
\begin{equation}
	\rho_{\rm sys}(t)-\rho_{\rm sys}(t-T_{\rm m})={\rm Tr}_{\rm bath}\left\{\int^{t}_{t-T_{\rm m}}ds[H_{\rm 
	int}(s),\chi(t-T_{\rm m})]\right\},
\end{equation}
where we have assumed that ${\rm Tr}_{\rm bath}\{H_{\rm 
int}(s)\rho_{\rm bath}(t_{0})\}=0$, as the mean bath interaction energy 
can always be taken inside the Hamiltonian for the system. This 
demonstrates that (even to first order 
in the perturbation) the change in the 
density matrix of the system at time $t$ is determined by the correlations that 
existed between the bath and the system at $t-T_{\rm m}$. Note that 
the master equation for the system for a Markov process is derived by 
going to second-order in the perturbation (correlations can only 
arise from an initial factorized density matrix by going to at least 
second order in the perturbation) and assuming that the correlations 
decay  instantaneously $T_{\rm m}\rightarrow 0$ on the time scale of the inverse 
coupling strength (see for example \cite{qn}).

Another consequence of the above mentioned entanglement is that to 
determine the probabilities for a particular measurement result at 
time $t$ we need to consider the evolution of the system from a time 
$t-T_{\rm m}$ in the past up until time $t$. In addition,  even
 if the initial state of the system is pure it  will not stay pure under 
evolution conditioned on continuous measurement of the output,
unlike the Markov case \cite{carmichael,wiseman}.
  
\begin{figure}
\begin{center}
\epsfig{file=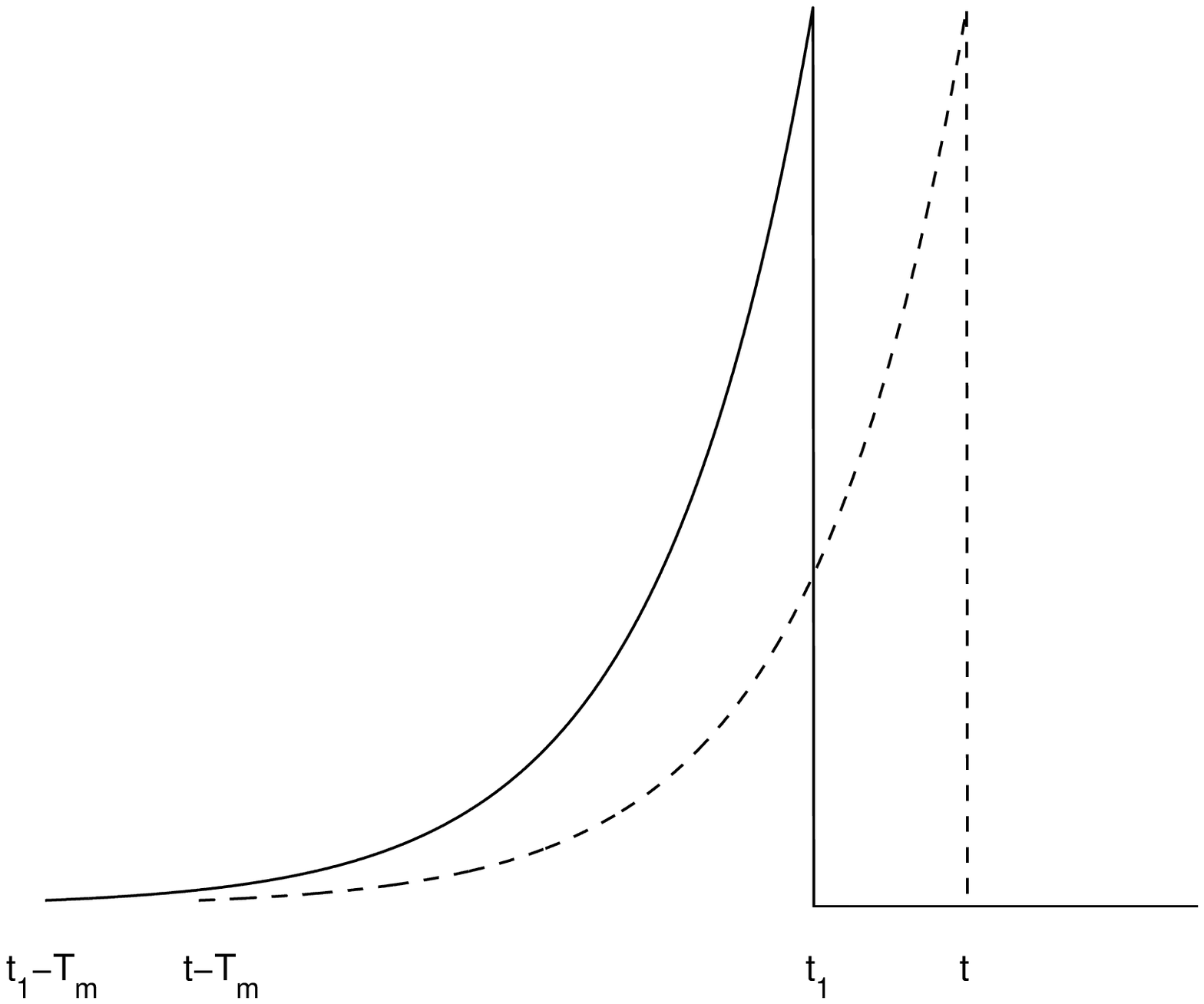,height=0.5\linewidth,width=0.6\linewidth}
\end{center}
\caption{\protect\footnotesize Diagram of the weighting	distribution
over the times of emission for a particle that may be detected in the
output at time $t$ (dashed line) and that has been detected	at time
$t_{1}$	(solid line). The times	at which previous emissions	can	be neglected
$t-T_{\rm m}$ and $t_{1}-T_{\rm	m}$	are	also shown.
\label{mixedstate}}
\end{figure}

Consider the simplest case of continuously counting  the emitted 
particles (e.g., a photodetector counting photons).  Suppose that a particle that is 
detected at time $t$  could only have been emitted during the 
time interval $[t,t-T_{\rm m})$, with a specific weighting for 
each possible emission time. Therefore, to determine the probability 
of a detection at time $t$ we need to propagate 
the system undergoing emissions over the time interval $[t,t-T_{\rm 
m})$. A diagram of this situation is given in Fig.\ref{mixedstate}, 
showing a possible weighting distribution for each emission time for a particle 
detected at time $t$. Now assume that we have a measurement 
record over the interval $[t,t-T_{\rm m})$ where a single particle was 
detected at time $t>t_{1}>t-T_{\rm m}$, (see Fig.\ref{mixedstate}). We have the complete measurement record 
from $t$ to $t-T_{\rm m}$. Therefore, the conditioned state of the emitting system at time 
$t-T_{\rm m}$ cannot change due to future detections. We are 
interested, then, in the conditioned state at time $t-T_{\rm m}$. 
The 
detection at time $t_{1}$ corresponds to two possibilities at time $t-T_{\rm m}$, either 
the particle was emitted after time $t-T_{\rm m}$ or before. In the 
Hilbert space of the system and bath this 
 corresponds to a superposition of the two possibilities: a particle 
has been as emitted and is present in the bath and a particle has not 
been emitted and the bath is empty. In terms of the system alone, when 
a trace over the bath is performed, this 
is a mixed state. Note that the mixed state is produced
by ignoring the state of the bath. As far as future evolution of the 
system is concerned
 there is no classical uncertainty in the 
state of the system at this time, each of the ket vectors in the 
density matrix  undergo a 
separate conditioned evolution, 
determined by the fact that the system must undergo one and only one 
emission during the interval $[t_{1},t_{1}-T_{\rm 
m})$ (and also by any future measurement results). 

It is a surprising fact that despite these complications it is still possible to derive quantum 
trajectories for non-Markovian processes. 

In order to deal with this sort of mixed conditioned state we generalize the 
notation
of Eq.(\ref{dotproduct}) and write 
\begin{equation}
	\Omega_{RI}(t,s)\circ|\tilde{\psi}(s)\rangle\equiv\sum_{k}\Omega_{RI}^{k}(t,s)
	|\tilde{\psi^{k}}(s)\rangle,
\end{equation}
where 
\begin{eqnarray}
	\Omega_{RI}^{k}(s) & = & \langle R(t,t_{0})|U_{\rm int}(t,s)|k\rangle,  \\
	|\tilde{\psi^{k}}(s)\rangle & = & \langle k|U_{\rm int}(s,t_{0})|I(t,t_{0})\rangle|\psi(t_{0})\rangle.
\end{eqnarray}
and the tilde denotes that the ket in unnormalized. 
Note that for any time $s$ there is always one state of the bath, 
 $|I(t,s)\rangle$, that is independent of 
the  measurement results. Let us then take this state as one of the basis states of 
the bath, $|k=0\rangle\equiv|I(t,s)\rangle$.

Having an explicit form for $\Omega$ means that we are able to 
calculate the 
probabilities for the results of measurements made on the output in 
terms 
of the conditioned system evolution alone. We will now relate this to 
 simulating the evolution of the measurement process forward in time. 
 To do this in any practical way a necessary requirement is that 
there 
 exists a time interval $T_{\rm m}$ such that a measurement made 
 at time $t$ will be independent of the state of the system before time 
 $t-T_{\rm m}$. In other words, there must exist a finite {\em memory-time} 
 for the open system. We will discuss this in more detail in 
Sec.\ref{memorytime}.

Consider a single run in which we have one particular input state and a record of the results of the 
measurements during the
 interval 
$[t,t_{0})$  and we would like to know the probability of a
 particular 
outcome during the next infinitesimal interval.
The conditional probability density of 
getting a particular result during the next infinitesimal interval 
$[t+dt,t)$ given 
the previous results over the interval $[t,t_{0})$ is
\begin{eqnarray}
P[ R(t+dt,t)| R(t,t_{0})I(t,t_{0})]   & = &  
\sum_{I(t+dt)}\frac{P[R(t+dt,t_{0})I(t+dt,t_{0})]}{P[R(t,t_{0})I(t,t_{0})]},\\
& = &  
\sum_{I(t+dt)}\frac{P[R(t+dt,t_{0})|I(t+dt,t_{0})]}{P[R(t,t_{0})|I(t,t_{0})]}
P[I(t+dt,t_{0})|I(t,t_{0})].
\end{eqnarray} 
If in a simulation we choose the input states consistent with 
$P[I(t+dt,t)|I(t,t_{0})]$ for each run then 
$P[R(t+dt,t_{0})|I(t+dt,t_{0})]/P[R(t,t_{0})|I(t,t_{0})]$ will 
generate results $R(t+dt,t)$  with the required probability distribution. 
These conditional probability densities can be written in terms of the 
operation as 
\begin{equation}
\frac{P[R(t+dt,t_{0})|I(t+dt,t_{0})]}{P[R(t,t_{0})|I(t,t_{0})]}
=\langle\tilde{\psi^{0}}(t+dt)|\tilde{\psi^{0}}(t+dt)\rangle,  
\end{equation} 
where we have defined the system state vector $|\tilde{\psi^{0}}(t+dt)\rangle$, 
at time $t+dt$, as
\begin{equation}
	|\tilde{\psi^{0}}(t+dt)\rangle=|\psi^{0}(t)\rangle+ 
	{\mathcal N}d\Omega_{R I}(t,t-T_{\rm m})\circ|\psi(t-T_{\rm 
m})\rangle,
	\label{probe}
\end{equation}
where $|\tilde{\psi^{0}}(t)\rangle=\Omega_{R I}(t,t-T_{\rm 
m})\circ|\psi(t-T_{\rm m})\rangle$ and $
{\mathcal N}=1/\sqrt{\langle\tilde{\psi^{0}}(t)|\tilde{\psi^{0}}(t)\rangle}$, 
again the tilde denotes that the 
state is unnormalized. 
This state determines the probabilities for the next measurement 
results. It is a transient state, 
in that it can be overwritten by the operation corresponding to 
subsequent 
measurement results (it is only the state of the system at time $t$ if 
the measurement results from $t$ to $t+T_{\rm m}$ correspond to a 
state equivalent to the 
input state during this time).
In Eq.(\ref{probe}) we have explicitly used the idea that the 
probabilities of 
getting a particular measurement result at time 
$t$ do not depend on the state of the system or bath before time 
$t-T_{\rm m}$. If this is true then the system state  at time $t-T_{\rm m}$ 
is then $\rho^{RI}_{\rm sys}(t-T_{\rm m})=
\sum_{k}P_{k}(t-T_{\rm m})|\psi^{k}(t-T_{\rm m})\rangle\langle\psi^{k}(t-T_{\rm m})|$
and is independent of the measurement 
results 
at time $t+dt$. Here, $P_{k}(t-T_{\rm m})={\mathcal M}\langle\tilde{\psi^{k}}(t-T_{\rm 
m})|\tilde{\psi^{k}}(t-T_{\rm m})\rangle$, where ${\mathcal M}$ is a 
constant which ensures that $\sum P_{k}(t-T_{\rm m})=1$. This conditioned state converges to the reduced density matrix 
for the system when averaged over many runs of the above trajectories,
 $\rho_{\rm sys}(t-T_{\rm m})=\sum_{IR}\rho^{IR}_{\rm sys}(t-T_{\rm m})$. In the following sections we will investigate 
the nature of $\Omega$ for a particle conserving coupling and some 
quite general input  states and measurement processes. 

\subsection{System-bath interaction} \label{interaction}

We assume that the interaction is linear and conserves 
particle number 
and also, for simplicity, that the bath is coupled to a 
single mode of the system. 
We consider situations that have bath and interaction Hamiltonians of 
the form
\begin{eqnarray}
	H_{\rm bath} &	= &	\int d{\bf k} \omega_{\bf k}b^{\dagger}({\bf k})b({\bf 
k}),\\
	H_{\rm int} &	= &	i\sqrt{\gamma}\int
	d{\bf k}\left\{g_{{\bf k}}b^{\dagger}({\bf 
	k})a-g^{*}_{{\bf
	k}}b({\bf k})a^{\dagger}\right\},
	\label{interaction hamiltonian}
\end{eqnarray}
where we have put $\hbar=1$ and $a$ and $b({\bf k})$ are the 
annihilation operators for the  modes of the system
and	bath. The bath modes satisfy Bose commutation relations 
$[b({\bf 
k}),b^{\dagger}({\bf k'})]=\delta({\bf k-k}')$. Note that our 
treatment does not depend on the system being a Bose field. The mode 
operator could equally well be replaced by,  for example, the lowering operator of a 
two-level atom.  The frequency 
of the ${\bf k}$th mode is $\omega_{\bf k}$ and is an arbitrary 
function 
of ${\bf k}$. The effective coupling constant, $g_{{\bf k}}$, is 
normalized so that 
$\int d{\bf k} |g_{{\bf k}}|^{2}=1$, and the 
strength of the coupling is given 
by $\sqrt{\gamma}$. The spatial dependence of the bath-system 
coupling 
determines $g_{{\bf k}}$,
\begin{equation}
	\sqrt{\gamma}g_{{\bf k}}=\int d{\bf r} 
\kappa({\bf 
	r})u^{*}_{\bf k}({\bf r}),
	\label{coupling constant}
\end{equation}
where $\kappa({\bf r})$ is the effective spatially dependent coupling 
constant
and $u_{\bf k}({\bf r})$ are the spatial 
modes of the  bath. An essential assumption 
in the 
following derivation is that the system and bath only interact over a 
finite
region $R$ localized about the origin so that we can put 
$\kappa({\bf r})\approx 0$ for $|{\bf r}|>R$. 
This is necessary so that we can later make the finite memory-time 
approximation.

We can define input and output bath fields \cite{hope,collett} outside the interaction 
region ($|{\bf r}|>R$) by
\begin{equation}
	b^{\rm in}(t,{\bf r})=\int d{\bf k} 
b({\bf k},t_{0})e^{-i\omega_{\bf k}(t-t_{0})}u_{\bf k}({\bf r}),
	\label{input}
\end{equation}
where $t_{0}<t$ can be taken as the initial time  and
\begin{equation}
	b^{\rm out}(t,{\bf r})=\int d{\bf k} 
b({\bf 
	k},t_{1})e^{-i\omega_{\bf k}(t-t_{1})}u_{\bf k}({\bf r}),
	\label{output}
\end{equation}
where $t_{1}>t$ can be taken as a time in the distant future.

The unitary evolution operator $U_{\rm int}(t,t_{0})$ is then given by the 
total Hamiltonian Eq.(\ref{total 
hamiltonian}) in the interaction picture,
\begin{equation}
	U_{\rm int}(t,t_{0})=T\exp\left\{i\sqrt{\gamma}\int^{t}_{t_{0}} 
	ds\left[\xi^{\dagger}(s)a_{I}(s)-\xi(s)a_{I}^{\dagger}(s)\right]\right\},
	\label{unitary interaction}
\end{equation}
where 
\begin{equation}
\xi(t)=\int d{\bf k} g_{{\bf k}}b({\bf k},t_{0})e^{-i\omega_{\bf 
k}(t-t_{0})},
\end{equation} 
is the driving field in the Langevin equations of motion for the 
system mode. $a_{I}(t)$ is the system mode operator in the 
interaction picture 
and $T$ denotes that the operators in the exponential are 
time-ordered.
We define a memory function $f_{\rm m}(t-t')$ as the commutation 
relation 
of the driving field with itself at an earlier time
 \begin{eqnarray}
 	f_{\rm m}(t-t') & \equiv & [\xi(t),\xi^{\dagger}(t')], 
 	 \label{memory function}\\
 	& = & \int d{\bf k} 
 	|g_{{\bf k}}|^{2}e^{-i \omega_{\bf 	k}(t-t')}.
 \end{eqnarray}

\subsection{Measurement channels}\label{channels}
The quantum trajectory formalism presented here is based on an 
idealized situation where 
all the output from the bath is eventually measured by a perfectly 
absorbing measuring device. The physical 
situations that can be modeled by this formalism are not limited by 
this, 
as the information acquired from any measuring device that is not 
really 
there in the physical situation can be averaged over a number of runs 
of the simulated measurement process. So this 
formalism, although based on measurement theory, contains as a 
special case 
even the situation when no measurements are being made at all and the 
system is 
simply decaying into the bath.

To describe separate measuring devices it is useful to introduce the 
concept of a complete set of measurement channels. A measurement 
channel has an associated 
field that contains only a range of modes of the original bath field. 
A measurement made on a particular channel is specific to the 
particular range of bath modes defined by the channel. The channels 
form a complete set, in that, together they contain the entire bath 
field. 

In a three-dimensional situation the 
bath output in all directions is being measured by detectors 
distributed around the system. To model 
the situation where each detector is sensitive only to the 
modes propagating towards it we can define channel fields containing 
only 
the appropriate bath modes for each detector. This concept of 
a channel also stretches to the case of spectral detection, i.e., 
measurements that directly access spectral information about the 
output 
field in a particular direction. A set of measurement channels can be 
defined such that each channel 
field contains only a range of frequencies of the output field. In 
this 
way, 
the results of a measurement made on one of these channel fields is 
specific to the 
particular range of frequencies defined by the channel.

As a by-product, this formalism allows for the consideration of 
measurement devices 
that have a non-trivial spectral response. We can do this by 
choosing one channel to have the required frequency response, keeping 
its measurement record, and averaging the measurement results of 
the other channels (that make up the complete set) over a number of 
runs.

In summary, we are considering a model where the output from the 
system passes through a 
generalized spectrometer, is scattered into different 
measurement channels and is then measured. The measurements are 
assumed to take place instantaneously at a single spatial point for 
each channel.
The theory as it stands does not model the situation where the 
particles can scatter off 
the measuring device and reinteract with the system.

\begin{figure}
\begin{center}
\epsfig{file=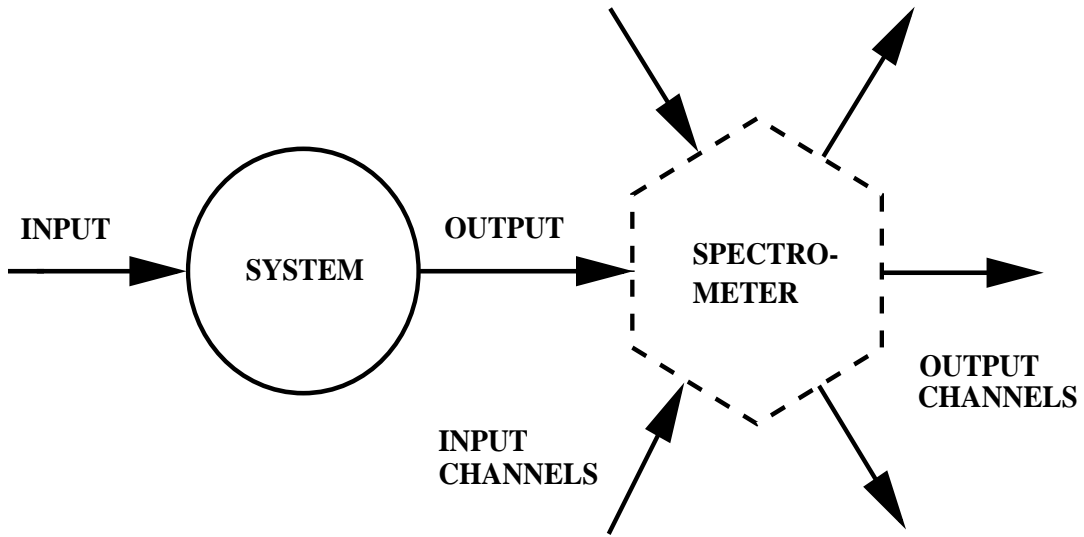,height=0.4\linewidth,width=0.8\linewidth}
\end{center}
\caption{\protect\footnotesize Schematic of	the	spectral detection
setup. The output from the system of interest is an	input to one of
the	spectrometer channels. The inputs to the other channels	are
simply the vacuum. The spectrometer	mixes the channel inputs so	that the
output channels	contain	certain	spectral components	of the original	system
output (as well	as contributions from the other	channel	inputs).
 \label{spectrometer}}
\end{figure}

The generalized linear-passive spectrometer is modeled by 
splitting the original output bath up into channels. A schematic of 
the situation is shown in Fig.\ref{spectrometer}. The number of input 
channels to the spectrometer must equal the number of output channels.
  
We define the $n$th channel output field by,
\begin{equation}
	c^{\rm out}_{n}(t,{\bf r})=\int d{\bf k}S_{n}({\bf k})
b({\bf k},t_{1})e^{-i\omega_{\bf k}(t-t_{1})}u_{\bf k}({\bf r})+ \mbox{other baths}.
	\label{channel output}
\end{equation}
where $S_{n}({\bf k})$ is the spectral 
response of the channel. The contributions from 
other free baths (with the same dispersion relation as $b({\bf k})$)  
do not interact with the 
system and are initially in a vacuum state. They do not concern us in 
latter derivations but 
they are 
necessary to preserve commutation relations (in other words they 
have a physical origin). 

The definition of the channel fields in Eq.(\ref{channel output}) is 
in fact a linear transform of 
the mode operators of the independent baths, $b_{m}(k)$, to the 
mode operators of the channels, $c_{n}(k)$, for each $k$ (we have 
written ${\bf k}$ as $k$ for ease of notation). Writing $b_{1}(k)=b(k)$ 
for 
the original bath modes and defining $c_{n}(k)$ by $c_{n}^{\rm in}(t,r)=\int dk 
c_{n}(k)e^{-i\omega_{k}(t-t_{0})}u_{k}(r)$,
we can write the transformation as
\begin{equation}
	c_{n}(k)=S_{n1}(k)b_{1}(k)+\sum_{m\neq 1}S_{nm}(k)b_{m}(k).
	\label{unitary transform}
\end{equation}
The requirement that the channel fields are independent and preserve commutation 
relations,
$[c_{n}(k),c^{\dagger}_{n'}(k')]=\delta_{nn'}\delta(k-k')$, is 
expressed as a constraint on  
$S_{nm}(k)$, 
\begin{equation}
	\sum_{m}S_{nm}(k)S^{*}_{mn'}(k)=\delta_{nn'}
\end{equation}
for all $k$. 
A special case of this when  $n=n'=1$ is
\begin{equation}
	\sum_{m}|S_{m}(k)|^{2}=1.
\end{equation}
where we have written $S_{n}(k)=S_{n1}(k)$. This expresses the 
fact that the channel fields are a complete decomposition of 
the bath field.

The $n$th output channel field  is given by a unitary 
transform  of the $n$th input 
channel 
field (the portion of the initial input bath that contributes to 
the measured output),
\begin{eqnarray}
	c_{n}^{\rm out}(t,{\bf r}) & = & U^{\dagger}_{\rm int}(t,t_{0})
	c^{\rm in}_{n}(t,{\bf r})U_{\rm int}(t,t_{0}),\\
	& = & \sqrt{\gamma}\int^{t}_{t_{0}} ds h_{n}(t-s,{\bf r})a(s)+
	c^{\rm in}_{n}(t,{\bf r}).
\end{eqnarray}
Note that the channel output fields contain a 
contribution from the system and the input channel fields.

We have defined  $c^{\rm in}_{n}(t,{\bf r})$ as,
\begin{equation}
	c^{\rm in}_{n}(t,{\bf r})=\int d{\bf k} S_{n}({\bf k})b({\bf k},t_{0})
e^{-i\omega_{\bf k}(t-t_{1})}u_{\bf k}({\bf r})+ \mbox{other baths}
	\label{channel input}
\end{equation}
and have also defined the impulse response functions, 
$h_{n}(t-s,{\bf r})$, for each channel, as the commutation relation of 
 $c^{\rm in}_{n}(t,{\bf r})$ with the driving field $\xi^{\dagger}(s)$ 
at an earlier time,
\begin{eqnarray}
	h_{n}(t-t',{\bf r}) & \equiv & [c^{\rm in}_{n}(t,{\bf r}),\xi^{\dagger}(t')],\label{impulse responce}\\
	& = &\int d{\bf k} S_{n}({\bf k})g_{\bf k}
	e^{-i\omega_{\bf k}(t-t')}u_{\bf k}({\bf r}) 
	\end{eqnarray}
where $t\geq t'$. As the $n$th channel field is always measured at the same point in space,
 ${\bf r}_{n}$, 
we introduce the notation $h_{n}(t-s)\equiv h_{n}(t-s,{\bf r}_{n})$. 
Definition Eq.(\ref{channel output}) yields the intuitively obvious relation 
with the output field (see Fig.\ref{spectrometer}),
\begin{equation}
c_{n}^{\rm out}(t,{\bf r})=\int^{t_{1}}_{t}dsh_{n}(t-s)b^{\rm out}(s,{\bf 
r}) +\mbox{other baths}.
\end{equation}

The functions $h_{n}(\tau)$ and $f_{\rm m}(\tau)$, defined by the 
commutation relations Eq.(\ref{impulse responce}) and
 Eq.(\ref{memory function}),  completely 
determine the characteristics of the system-bath interaction and 
measurement process. In the next subsection we will consider some 
general properties of these functions.

\subsection{Finite memory-time} \label{memorytime}
We now introduce an essential concept, that of the existence of a finite memory-time. 
That is, we 
assume that both the memory function and the impulse responses are 
such 
that at some finite time in the past $T_{\rm m}$ the following 
conditions hold
\begin{eqnarray}
\left|\int^{t}_{t-T_{\rm m}} dsf_{\rm m}(t-s)a(s)\right| & \gg &
\left|\int^{t-T_{\rm m}}_{t_{0}} dsf_{\rm m}(t-s)a(s)\right|, 
\label{memfunfinite}\\
\left|\int^{t}_{t-T_{\rm m}} dsh_{n}(t-s)a(s)\right| & \gg &
\left|\int^{t-T_{\rm m}}_{t_{0}} dsh_{n}(t-s)a(s)\right|,
\label{imprespfinite}\end{eqnarray}
for all $n$ and all $t$ during the entire measurement interval.  To 
guarantee the above condition we assume that the memory and 
impulse functions satisfy 
\begin{eqnarray}
\int^{t}_{t-T_{\rm m}} ds\left|f_{\rm m}(t-s)\right|^{2} & \gg &
\int^{t-T_{\rm m}}_{t_{0}} ds\left|f_{\rm m}(t-s)\right|^{2}, \\
\int^{t}_{t-T_{\rm m}} ds\left|h_{n}(t-s)\right|^{2} & \gg &
\int^{t-T_{\rm m}}_{t_{0}} ds\left|h_{n}(t-s)\right|^{2},
\end{eqnarray}  
which are independent of $a(s)$. These are useful working conditions 
and are satisfied by all the functions considered in this paper. 
However, it is also possible to satisfy Eq.(\ref{memfunfinite}) and 
Eq. (\ref{imprespfinite})
 if the memory function an impulse responses have a rapidly oscillating behavior.  In 
 this case the contribution to the integral after some time 
 $t-T_{\rm m}$ may average to zero.

To understand the consequences of a finite memory-time we 
consider the physical effect of the driving field on the 
system. The Heisenberg equation of 
motion for $a(t)$ is
\begin{equation}
	\frac{da(t)}{dt}\approx -i[H_{\rm 
	sys},a(t)]-\frac{\gamma}{2}\int^{t}_{t-T_{\rm m}}f_{\rm 
	m}(t-s)a(s)+\xi(t),
	\label{heisenberg}
\end{equation}
where we have neglected the smaller part of the integral.
Correlations of the driving field with itself at earlier times 
cause the evolution of 
the system to be dependent on its past behavior. The existence of a memory-time 
implies that there is a finite time after which one can 
neglect the effect of the past state of the system in determining the 
present evolution. 

The physical effect of the system on the output channel field is given by 
\begin{equation}
		c_{n}^{\rm out}(t,{\bf r}) \approx \sqrt{\gamma}\int^{t}_{t-T_{\rm m}} ds 
		h_{n}(t-s,{\bf r})a(s)+c^{\rm in}_{n}(t,{\bf r}),
	\label{input-output}
\end{equation}
where we have again explicitly inserted the lower bound of the 
integral. From this equation one can see that the state of the system 
in the past determines the output channel field in the present and 
that the finite memory time imposes a cutoff time for this dependence. 
This  relation highlights an important point 
 regarding the interpretation of the results of measurements
  of the output channels in terms of the system dynamics. 
  A measurement result at time $t$ does not mean that we can ascribe a
   state corresponding to this 
 result to the system at the same time $t$. Unmeasured output at later 
 times is still quantum mechanically entangled with the system at time $t$ 
 and a subsequent measurement could require us to revise our knowledge of the system
   state completely. In the Markov case, because of the delta function
    impulse response, a 
 measurement result meant that we could specify the state of the
  system at the same time. In the general case considered here, only
   states of the system at a time
    $t-T_{\rm m}$ can be thought of as fully specified, as the output
     after time $t$ is uncorrelated with the system before this time 
     and future measurements will not alter our knowledge of the 
     system state.  
We can then  interpret the state at time $t-T_{\rm m}$ as the system state 
conditioned on the measurements up to and including time $t$.

\subsection{Conditioned system evolution}
We now return to a treatment in terms of state vectors and set about 
determining the explicit form of $\Omega_{RI}$ for a general 
measurement scheme. 

 We consider quantum 
trajectories to be an attempt to simulate the measurement results of 
a distant measuring device by the evolution of a local system emitting 
particles. This is related to the way in which the spectrometer is 
treated in this work. Instead of explicitly modeling the generalized 
spectrometer 
as a physical system (coupled to our localized system) 
we move towards a signal processing treatment, where the spectrometer 
is treated as a black box and its inputs are related to its outputs 
via a spectral response.

Take, for example, a number of measuring devices making particle counting 
measurements. At any one time we assume that there are only two 
possible outcomes (for perfect detectors): the null result, where no 
particle is present at 
any of the detectors; 
and the positive result, where a particle is detected at a single 
detector. 
A positive result corresponds, in general, to a superposition of 
particle emissions from the system at times in the past. Therefore, a 
probability can be calculated for the event from only the evolution 
of the system undergoing emissions. We assume that the weighting given to each emission time 
characterizes completely the passage of the particle to the detectors. 
The 
probability of the null event can be inferred from the requirement 
that 
the probabilities of all possible measurement results must add up to one. 
The probability of the null event 
cannot be determined directly as we are not explicitly modeling the 
physical 
evolution of the particle from the system to the detector.

It is a rather interesting fact that in the spectral detection case the 
null result, the no detection 
result, cannot be assigned an operation.  In the 
Markovian quantum trajectory case a null result corresponds 
precisely to a non-emission event (a virtual emission and absorption) 
and so an operation is also assigned to the null event. The virtual 
emission-absorption event is localized in the interaction region and 
so is not directly related to results at the distant detector. 
However, in the Markovian case the detector can in principle be 
brought right 
up against the interaction region (which takes place at a point in 
the 
Markovian case).

Recall that we have defined the operation in the system space corresponding 
to the 
measurement results over the interval $[t,t_{0})$ as 
\begin{equation}
 \Omega_{RI}(t,t_{0})=\langle { 
 	R}(t,t_{0})|U_{\rm int}(t,t_{0})|{I}(t,t_{0})\rangle,
\end{equation}
where $U_{\rm int}(t,t_{0})$ is given by Eq.(\ref{unitary interaction}). We are trying to simulate the measurement process 
in terms of the evolution of the system alone and so have not 
included 
the spectrometer in the unitary evolution. The operation contains all 
the information about how to evolve the system forward conditioned on 
measurement results and also how to calculate the probability of a 
particular outcome. 

We assume that our idealized measuring devices yield a result for 
each 
infinitesimal interval of the continuous measurement process. The 
state corresponding to the results must then factorize into a product 
of states over each infinitesimal interval. We can therefore write 
the output state as a product 
of state generators over each infinitesimal interval acting on the 
vacuum. We consider measurements where the generator for the output 
is written as  a function of the creation operators for the 
free 
input channel fields at a particular position $c_{n}^{\dagger {\rm 
in}}(t,{\bf r}_{n})$, (the measurement apparatus can be thought of as 
a sink for 
particles at position ${\bf r}_{n}$).
In  mathematical 
language we have,
\begin{equation}
	|{R}(t,t_{0})\rangle =  \prod_{n,j}G^{R}_{j,n}[c^{\dagger{\rm in}}_{n}(t_{j},{\bf r}_{n})]|\{0\}\rangle,
	\label{output generator}
\end{equation}
where $G^{R}_{j,n}[c_{n}^{\dagger {\rm in}}(t_{j},{\bf r}_{n})]$ is the generator of the 
state during the interval
 $[t_{j}+dt,t_{j})$ where $t_{j}$ ranges 
from $t_{0}$ to $t-dt$. The input channel operator appears in this 
expression because it is the interaction picture output channel field. For example, 
these generators may correspond 
to measurements of the quadrature 
amplitudes (homodyne \cite{carmichael} and heterodyne detection 
\cite{wisemanhetrodyne}) or  absorption measurements 
 (particle counting)\cite{carmichael}.

The input can similarly be thought of as a source of particles at 
positions ${\bf q}$ and is  written in terms of 
the input field, i.e.,
\begin{equation}
	|{I}(t,t_{0})\rangle =  \prod_{j}G^{I}_{j}[b^{\dagger{\rm in}}(t_{j},
	{\bf q})]|\{0\}\rangle.
	\label{input generator}
\end{equation}
For example, these generators can correspond to vacuum or coherent input states, 
squeezed input states or thermal input states \cite{dalibar,gardiner}. 

Recall that our problem is to determine the operation $\Omega_{RI}(t,t_{0})$ 
in terms of the system 
variables only.
Because of the factorized form of the input and output state we can 
consider operations of the form
\begin{equation}
O_{n}(t_{j})=\langle\{0\}|G_{j,n}^{R}[c_{n}^{\rm in}(t_{j},
{\bf r}_{n})]U_{\rm int}(t,t_{0})G_{j}^{I}[b^{\dagger{\rm in}}
(t_{j},{\bf q})]|\{0\}\rangle,
\end{equation}
for each $t_{j}$ and each channel $n$.
 
To proceed further we note that we can write the interaction unitary 
evolution 
operator $U_{\rm int}(t,t_{0})$ in the form 
\begin{equation}
	U_{\rm int}(t,t_{0})=T_{a}\left\{V_{+}V_{0}V_{-}\right\},
	\label{normalandtimeordering}
\end{equation}
where $T_{a}$ denotes time-ordering of the system operators $a_{I}$ 
and 
$a_{I}^{\dagger}$ and where 
\begin{eqnarray}
V_{+} & = & \exp\left\{\sqrt{\gamma}\int^{t}_{t_{0}}ds
 \xi^{\dagger}(s)a_{I}(s)\right\},\\
V_{0} & = &
\exp\left\{-\frac{\gamma}{2}\int^{t}_{t_{0}}ds_{2}\int^{s_{2}}_{t_{0}}ds_{1}
f_{\rm m}(s_{2}-s_{1}) a_{I}^\dagger(s_{2})a_{I}(s_{1})\right\},\\
V_{-} & = &\exp\left\{-\sqrt{\gamma}\int^{t}_{t_{0}}ds 
\xi(s)a^\dagger_{I}(s)\right\}.
\end{eqnarray}
This can be proved by differentiating $U_{\rm int}(t,t_{0})$ and 
re-ordering 
the bath operators.
The operation $O_{n}(t_{j})$ then becomes
\begin{equation}
	O_{n}(t_{j})=\langle \{0\}|T_{a}\left\{G^{R}_{j,n}[c^{\rm 
	in}_{n}(t_{j},{\bf 
r}_{n})]V_{+}V_{0}V_{-}G_{j}^{I}[b^{\dagger{\rm 
in}}(t_{j},{\bf q})]\right\}|\{0\}\rangle,
\end{equation}
where we have expanded the brackets of the time-ordering to encompass 
the state generators as they are only a function of the bath 
operators.

We now make use of the relation (compare with Eq.(\ref{channel 
output}))
\begin{eqnarray}
	V_{-}b^{\dagger{\rm in}}(t,{\bf q})V_{-}^{-1} & = & 
	b^{\dagger{\rm 
	in}}(t,{\bf q})+\sqrt{\gamma}{\mathcal 
A}^{\dagger}(t,t_{0};s), \\
	V_{+}^{-1}c^{\rm in}_{n}(t,{\bf r}_{n})V_{+} & = & 
	c^{\rm in}_{n}(t,{\bf r}_{n})+\sqrt{\gamma}{\mathcal 
A}_{n}(t,t_{0};s)
\end{eqnarray}
where we have defined for notational purposes 
\begin{eqnarray}
	{\mathcal A}_{n}(t,t_{0};s) & \equiv & \int^{t}_{t_{0}} ds 
	h_{n}(t-s)a_{I}(s),\\
{\mathcal A}(t,t_{0};s) & \equiv & \int^{t}_{t_{0}} ds 
	[b^{\rm in}(t,{\bf q}),\xi^{\dagger}(s)]a_{I}(s).
\end{eqnarray}

We then make the above replacements and note that 
$V_{-}$ can be annihilated on the 
right-hand vacuum and $V_{+}$  on the left-hand vacuum. The the 
system operators $a_{I}$ and 
$a_{I}^{\dagger}$ commute with all the input fields 
$c^{\rm in}_{n}(t_{j},{\bf r}_{n})$ and $b^{\rm 
in}(t_{j},{\bf q})$, as they are still inside the 
time-ordering brackets and so act like complex numbers. 

If we repeat this procedure for each time $t_{j}$ and if we restrict 
the nature of the state generators so 
that when the $R$ and $I$ generators for each time are swapped from 
left to right 
no bath modes are left that are not normally-ordered (specific 
examples will be given latter) then we are left with an operation of 
the form
\begin{equation}
		\Omega_{RI}(t,t_{0})=T\left\{\prod_{j,n}G_{j,n}^{R}[\sqrt{\gamma}{\mathcal A}_{n}(t_{j},t_{0};s)]
	V_{0}\prod_{j}G^{I}_{j}[\sqrt{\gamma}{\mathcal 
	A}^{\dagger}(t_{j},t_{0};s)]\right\},
	\label{finaloperation}
\end{equation}
where we have dropped the reference to the mode operators in the time 
ordering operator $T$.

This is the general result for arbitrary (with the above 
restriction) measurement schemes and arbitrary input. No assumptions 
have been made regarding the strength of the coupling constant or the 
nature of the correlations between the system and the bath after the 
initial time. This operation can be transformed to the 
Schr\"{o}dinger picture by multiplying on the left by the 
unitary evolution operator for the free system.

Because of the complex nature of the above operation the measurement 
process can only be simulated on a computer in the weak-interaction 
short-memory-time case where 
$I(t)/\Gamma<1$,  for all times of interest, where
 $I(t)=\langle b^{\dagger{\rm out}}(t)b^{\rm 
out}(t)\rangle\sim\gamma\langle a^{\dagger}(t)a(t)\rangle$ is the intensity 
of the output. $\Gamma\sim 1/T_{\rm m}$ is some measure of the rate of 
decay of the of the memory function or impulse response. It determines, in essence, 
 how non-Markovian the process is. In this situation  we can make a 
perturbative expansion of the operation $\Omega_{RI}$
 in powers of $\gamma\langle a^{\dagger}(t)a(t)\rangle/\Gamma$. To lowest 
orders the dimensionality of the 
integrals during a memory-time will be small and  they can be evaluated 
numerically  with a computer. In absorption measurements each 
click of the counter corresponds to one integral so this can be simulated in the 
case where there are on average only a few detections (real emissions) 
during a memory-time (there will also be integrals of the same 
dimension 
corresponding to 
virtual emission-absorption events).  We shall explicitly demonstrate how this can be 
done 
in the case of optical spectral detection in the next section.

\section{Optical spectral detection}\label{two}
In this section we consider a special case of the above theory; that 
of spectral detection of an optical field emitted from a source. 
To check the accuracy of our numerical simulations we shall apply it to a known 
problem; the spectral measurement of light emitted from a 
strongly driven two-level atom. 

Spectral detection of an optical field is defined here as meaning 
that each 
photodetection can be identified with the emission of a photon from a 
certain part of the source spectrum \cite{wisemanthesis} (this is in 
contrast to measurements 
from which the spectrum can be computed, such as heterodyne 
detection). 
Spectral detection necessarily involves the interference of light 
emerging 
from the system over some time period $\Delta t$. The length of this 
time 
period varies inversely with the accuracy of the spectral measurement. 

The type of spectral detection that has been considered previous to 
this 
work is that of detecting fields with spectral peaks of widely 
differing frequencies.
These are then just treated as independent Markovian baths for the 
source to decay into. 
The evolution of the wave function for this type of system can then be
treated with the standard quantum trajectory approach.  Here we would 
like to go beyond this and treat wave function 
evolution for the source conditioned on spectral detection of emitted 
photons 
{\em within} an emission linewidth. To achieve this it is necessary to give up knowledge
of the exact time of emission. We can only know the time of emission 
to within an interval
$\Delta t>1/\Delta \omega$ if we are to resolve the emission spectrum 
to 
within $\Delta\omega$ of a certain frequency.

\subsection{The spectrum of a strongly driven two level atom}
A strongly driven two-level atom  can be described (in a frame rotating at 
the frequency of 
the driving field) by the
 free system Hamiltonian,
\begin{equation}
	H_{\rm atom}=\frac{\Omega}{2}(\sigma^{\dagger}+\sigma),
	\label{atom}
\end{equation}
where $\sigma$ is the lowering operator and $\Omega$ is the classical 
driving field 
strength and we have assumed that the driving 
field is on resonance with the atomic transition. 

The optical field interacts with the atom in a very small region about ${\bf r}=0$ and we can write the unitary 
evolution 
Eq.(\ref{unitary interaction}) as
\begin{equation}
	U_{\rm int}(t,t_{0})=T\exp\left\{i\sqrt{\gamma}\int^{t}_{t_{0}}ds\left[\sigma^{\dagger}_{I}(s)\xi(s)
-\sigma_{I}(s) \xi(s)\right]\right\},
	\label{two level atom interaction}
\end{equation}
where in this case
\begin{equation}
	\xi(t)=\int^{\infty}_{-\infty}d\omega 
b(\omega,t_{0})e^{-i\omega(t-t_{0})}.
\end{equation}
We have replaced the mode label ${\bf k}$ by the angular frequency 
$\omega$ as 
is customary in optics in the situation when the direction of the 
emitted light is unimportant. We have also ignored the polarization 
of the light field. Note that $\xi(t)$ is the same as the free input 
field at the point of interaction ${\bf r}=0$ and it is delta-correlated with itself,
$[\xi(t),\xi^{\dagger}(t')]=\delta(t-t')$. Because of this delta 
function
correlation of the driving field a Markovian master equation can be 
derived for the system by tracing 
over the bath (see \cite{qn} for example). Here we will consider a non-Markovian {\em 
unraveling} 
\cite{carmichael} of 
this master equation into quantum trajectories . 

It has been shown by Mollow \cite{mollow} that the spectrum of the 
spontaneous emission from a strongly driven two-level atom has three 
peaks at 
the frequencies $0$, $\pm \Omega$. It has also been shown by 
Cohen-Tannoudji and Reynaud \cite{cohen} that the side-bands exhibit 
photon antibunching 
independently, but are correlated via 
photon bunching. Experiments by Aspect {\em et al}. \cite{aspect} have 
confirmed these 
theoretical results by frequency filtering the light 
incident on three photodetectors, so that each photodetector is  
only sensitive to a range of frequencies about one of the peaks of 
the 
output light. Here we would like to reproduce these well known results by simulating this 
``spectral detection'' of the 
spontaneously emitted light with a non-Markovian 
quantum trajectory for the source atom. 

The original theoretical predictions were made with a 
fully quantized 
atom-field model and the model was analyzed in terms of the 
eigenstates of the Hamiltonian, the dressed states of the atom. 
Instead, 
we will consider the simplified model described by the 
Hamiltonian above, Eq.(\ref{atom}). In the limit of a large classical 
driving the field can be assumed to be in a coherent 
state and our model is valid. All the important results of the 
dressed 
state model will be present in our treatment \cite{wisemanthesis}. 
We can qualitatively understand the behavior of this system by 
considering the evolution of the  interaction-picture lowering operator,
\begin{equation}
	\sigma_{I}(t)=\textstyle{\frac{1}{2}}
	\left(\sigma_{x}+|+\rangle\langle -|e^{i\Omega 
	t}+|-\rangle\langle+|e^{-i\Omega t}\right)
\end{equation}
where $|+\rangle,|-\rangle$ are 
the eigenstates of $\sigma_{x}=\sigma_{+}+\sigma_{-}$, $\sigma_{x}|\pm\rangle=\pm|\pm\rangle$.
If we are continuously monitoring the emission from the system and
 are able to resolve the frequencies of the emitted photons, then 
if a photon is detected in the $+\Omega (-\Omega)$ frequency peak the system will 
be projected into the opposite eigenstate $|-\rangle (|+\rangle)$. If, 
however, a photon is detected in the central or zero frequency peak 
the value of $\langle\sigma _{x}(t)\rangle$ will be unchanged. The 
interesting behavior of this system stems from this simple idea.

In the following sections we will only consider the case 
of direct 
detection, that is, absorption measurements by photodetection.
The state corresponding to an absorption measurement in one of the 
spectral channels during an 
infinitesimal interval is 
\begin{equation}
	|{R}(t+dt,t)]\rangle=\prod_{n}c_{n}^{\dagger{\rm 
	in}}(t)^{dN_{n}(t)}|\{0\}\rangle,
	\label{output state direct detection}
\end{equation}
where the restrictions $(\sum_{n}dN_{n}(t))^{2}=\sum_{n}dN_{n}(t)$ 
and 
$dN_{n}(t)^{2}=dN_{n}(t)$ describe a point 
process, such that in an infinitesimal interval, either only 
one photon is detected in one of the channels or 
no photons are detected in any of the channels. The $dN_{n}(t)$ 
can take the value $1$ or $0$ corresponding to a click of the 
photodetector for channel $n$  and no click of 
the photodetector. We also assume that the input 
state is the vacuum (the coherent driving being included in the 
system Hamiltonian). 

Consider the situation where $N$ photons have been detected at each 
of the 
times $t_{1}<t_{2}<\cdots<t_{i}<\cdots<t_{N}$ in the channels 
$n_{1},n_{2},\ldots,n_{i},\ldots,n_{N}$ during the past interval 
$[t,t-T_{\rm m})$. 
The wavefunction Eq.(\ref{probe})for the 
next infinitesimal interval is
\begin{eqnarray}
	|\tilde{\psi^{0}}(t+dt)\rangle & = &
	 |\psi^{0}(t)\rangle-dt\frac{\gamma}{2}\sigma_{I}^{\dagger}(t)\sigma_{I}(t)|\psi^{0}(t)\rangle\nonumber\\
	& +&\sum_{n}dN_{n}(t)\left[{\mathcal N}
	\sqrt{\gamma}\sum^{2^{N}}_{i=1}\int^{t}_{t-T_{\rm m}}ds 
	h_{n}(t-s)T\left\{\sigma_{I}(s)\Lambda^{i}(t,t-T_{\rm 
	m})\right\}|\psi^{i}(t-T_{\rm 
	m})\rangle-|\psi^{0}(t)\rangle\right],
	\label{probeforphotodetection}
\end{eqnarray}
where 
\begin{equation}
	\Lambda^{i}(t,t-T_{\rm m})=\prod\sqrt{\gamma}{\mathcal 
	A}_{n_{j}}(t_{j},t-T_{\rm m};s_{j})V_{0}(t,t-T_{\rm m}),
	\label{lambda}
\end{equation}
and 
 \begin{equation}
	{\mathcal A}_{n_{j}}(t_{j},t-T_{\rm 
m};s)\equiv\int^{t_{j}}_{t-T_{\rm m}} ds h_{n_{j}}(t_{j}-s)\sigma_{I}(s).
	\label{scriptAsigma}
\end{equation}
In this case $V_{0}(t,t-T_{\rm m})=\exp\{-\frac{\gamma}{2}\int^{t}_{t-T_{\rm 
m}}ds a_{I}^{\dagger}(s)a_{I}(s)\}$, this leads to a much  
simpler evolution than the general case. The integrals ${\mathcal A}_{n_{j}}(t_{j},t-T_{\rm m};s)$ have the 
lower bound $t-T_{\rm m}$. However, there is no factorization 
of the operation at $t-T_{\rm m}$. Instead, we must write a different 
state vector for each $\Lambda^{i}$. The product is over each 
combination of from $0$ to 
$N$ integrals. To understand what we have done here it is necessary to look at the 
integrals more carefully. Consider, as an example, two time-ordered 
integrals,
\begin{equation}
 T\{{\mathcal A}_{n_{1}}(t_{1},t_{1}-T_{\rm 
m},s_{1}){\mathcal A}_{n_{2}}(t_{2},t_{2}-T_{\rm m},s_{2})V^{0}(t,t_{2}-T_{\rm 
m})\}=\int^{t_{1}}_{t_{1}-T_{\rm m}}\int^{t_{2}}_{t_{2}-T_{\rm 
m}}T\left\{\cdots \right\},
\end{equation}
where 
$t>t_{1},t_{2}>t-T_{\rm m}$. For each detection at time 
$t_{i}$ there is 
an integration over emission times during the interval $[t_{i},t_{i}-T_{\rm m})$.
If we split both integrals into the sum of two parts - emissions 
before $t-T_{\rm m}$ and 
emissions after - we can write this as 
the sum of four terms
 \begin{eqnarray}
 \int^{t_{1}}_{t_{1}-T_{\rm m}}\int^{t_{2}}_{t_{2}-T_{\rm 
m}}T\left\{\cdots \right\} & = & \int^{t_{1}}_{t-T_{\rm m}}\int^{t_{2}}_{t-T_{\rm 
m}}T\left\{\cdots \right\}T\left\{\cdots \right\}  +  \int^{t_{1}}_{t-T_{\rm m}}T\left\{\cdots \right\}\int^{t-T_{\rm m}}_{t_{2}-T_{\rm 
m}}T\left\{\cdots\right\}\nonumber\\  
& + &  \int^{t_{2}}_{t-T_{\rm m}}T\left\{\cdots \right\}\int^{t-T_{\rm m}}_{t_{1}-T_{\rm 
m}}T\left\{\cdots\right\} 
  +   T\left\{\cdots \right\}\int^{t-T_{\rm m}}_{t_{1}-T_{\rm m}}\int^{t-T_{\rm m}}_{t_{2}-T_{\rm 
m}}T\left\{\cdots \right\}.
\end{eqnarray} 
This can be easily generalized to $N$ integrals. This is the essence 
of what has been done in Eq.(\ref{probeforphotodetection}). The parts 
of the integrals before $t-T_{\rm m}$ get absorbed into wave 
functions. Each wavefunction is labeled so that its subsequent 
evolution is consistent with its previous evolution. Physically this 
means that if a photon is detected in a channel at a particular time 
this corresponds to a sum over {\em individual} emission events and a photon 
cannot be emitted more than once. This is the detailed structure 
glossed over in the $\circ$-product notation of the previous section.  

The probability density for a detection to occur in channel $n$ 
during the 
time interval
$[t+dt,t)$ conditioned on the previous detections is then given 
by
\begin{equation}
	P_{n}(t+dt|t_{N},t_{N-1},\cdots, 
	t_{1})=\langle\tilde{\psi_{n}^{0}}(t+dt)|\tilde{\psi_{n}^{0}}(t+dt)\rangle,
\end{equation}	
	where
\begin{equation}
	|\tilde{\psi_{n}^{0}}(t+dt)\rangle={\mathcal 
N}\sqrt{\gamma}\sum^{2^{N}}_{i=1}\int^{t}_{t-T_{\rm m}}ds
	h_{n}(t-s)T\left\{a_{I}(s)\Lambda^{i}(t_{i},t_{i}-T_{\rm 
	m})\right\}|\psi^{i}(t-T_{\rm 
	m})\rangle.\label{psin}
\end{equation}

This result is of limited practical use in the above form
as evaluating  multi-dimensional integrals on a computer is a 
slow 
business. Instead we take a less ambitious approach and assume that 
the bath-system coupling is weak so that we can make what amounts to 
a perturbative expansion of the trajectory in terms of the coupling 
strength $\sqrt{\gamma}$. In other words, we assume that on 
average only 
a small number of detections occur during the time interval 
$[t,t-T_{\rm m})$, the number of detections determining the 
dimension of the integration. To first order, i.e., when the 
maximum 
number of detections per memory time is one, the trajectory is very 
similar to the Markovian case. We therefore consider the next highest 
order so 
that there is a finite possibility of two detections occurring per 
memory time.

In order to demonstrate how a trajectory is simulated on a computer 
in 
practice, we write down as an example the Schr\"{o}dinger picture form 
of 
Eq.(\ref{psin}) 
in the case where one detection occurred at time $t'$ in channel $n'$ 
during the 
previous memory time,
\begin{eqnarray}
	\lefteqn{\frac{|\tilde{\psi^{0}_{n}}(t+dt)\rangle}{\mathcal N}  =  
	\gamma\int^{t}_{t'}\! ds_{1}\int^{t'}_{t-T_{\rm 
	m}}\!\!\! ds_{2}h_{n}(t-s_{1})h_{n'}(t'-s_{2})U_{\rm 
	eff}(t,s_{1})\sigma U_{\rm eff}(s_{1},s_{2})\sigma U_{\rm 
	eff}(s_{2},t-T_{\rm m})|\psi^{0}(t-T_{\rm m})\rangle} \nonumber\\
	&  &+ \gamma U_{\rm eff}(t,t')\int^{t'}_{t-T_{\rm 
	m}}\!\!\! ds_{1}\int^{s_{1}}_{t-T_{\rm m}}\!\!\! ds_{2}
	[h_{n}(t-s_{1})h_{n'}(t'-s_{2})]_{\rm sym}
	U_{\rm eff}(t',s_{1})\sigma U_{\rm eff}(s_{1},s_{2})\sigma U_{\rm 
	eff}(s_{2},t-T_{\rm m})|\psi^{0}(t-T_{\rm m})\rangle \nonumber\\
	& &+  \sqrt{\gamma}\int^{t}_{t-T_{\rm m}}\!\!\! ds h_{n}(t-s)U_{\rm 
eff}(t,s)\sigma U_{\rm eff}(s,t-T_{\rm m})|\psi^{1}(t-T_{\rm m})\rangle.
	\label{schrodingerpicture}
\end{eqnarray}
where $U_{\rm eff}(t,t_{0})=\exp\{[-iH_{\rm 
atom}-\frac{\gamma}{2}\sigma^{\dagger}\sigma ](t-t_{0})\}$ and 
$[h_{n}(t-s_{1})h_{n'}(t'-s_{2})]_{\rm sym}=
h_{n}(t-s_{1})h_{n'}(t'-s_{2})+h_{n}(t-s_{2})h_{n'}(t'-s_{1})$. The 
right-hand 
side is the sum of three terms. The first term represents the situation 
where the emission corresponding to the detection at $t'$ occurs 
after 
time $t-T_{\rm m}$ but before the emission that is detected at time 
$t$. The second term 
represents the situation where both the $t$ emission and the $t'$ 
emission occur during the same time interval $[t',t-T_{\rm m})$ and 
the 
time-ordering of the emission is accounted for by the symmetric 
ordering of the  impulse response functions. In the last term the 
time $t'$ emission has occurred before time $t-T_{\rm m}$ (and is
 included in $|\psi^{1}(t-T_{\rm m})\rangle$)  and only 
the time $t$ emission is left. The states 
$|\psi^{i}(t-T_{\rm m})\rangle$ are propagated forward one time 
step $dt$ by doing the integrations contained 
in the operation from $t-T_{\rm m}$ to $t-T_{\rm m}+dt$. 

An important technical aspect of the simulation is; how to choose a 
finite length impulse 
response for numerical calculations given a particular spectral 
response?
This is a problem of signal analysis and numerical routines 
exist for automatically generating finite length impulse responses 
from a given spectral response such that the spectral response 
corresponding to the finite length impulse response is as close as 
possible  
to the original in a least-squares min-max sense \cite{digital}. Note 
that if the impulse response is too long there is a greater chance of 
having 
to deal with higher-order integrals, while if it is too short the spectral 
response will not approximate the original spectral response and the 
probabilities will not be accurate. In line with this we have avoided 
triple integrals 
by shortening the response functions  when calculating the 
probabilities for detection after two detections have already 
occurred 
during the previous memory-time. As the probability of three 
detections
 during a memory time is already small (in the limit of weak damping) 
this should 
not have a large effect on the statistics of the counts.

The simulation procedure at each time step $dt$ is as follows. First, 
the 
conditional probabilities  for 
detections at time $t+dt$ in each of the channels during the time 
step is generated by calculating the 
wave function Eq.(\ref{schrodingerpicture}) for each $n$. This depends 
on any previous detections during the time interval. The normalization 
${\mathcal N}=\langle\tilde{\psi^{0}}(t)|\tilde{\psi^{0}}(t)\rangle$ must also be 
calculated independently at each time step by propagating from 
$t-T_{\rm m}$ to $t$ conditioned on previous detections but assuming 
that there will be no more detections during the time interval $[t+T_{\rm 
m},t)$. The 
probabilities for detections in each channel are added together and compared to a random number. If 
the probability of a count is less than the random 
number, the state 
at time $t-T_{\rm m}$, $|\psi^{(i)}(t-T_{\rm m})\rangle$ is 
propagated forward conditioned on the fact that there was no 
detection at time $t+dt$. If the probability of a count was greater 
than the random 
number a channel is randomly picked from the probability distribution 
over the channels. The state at time $t-T_{\rm m}$ is then propagated 
forward $dt$ conditioned on this outcome. The process can then begin 
again.

\subsubsection{The Frequency Filter}
Here we consider the situation where we frequency filter the output 
from 
the atomic source. The light emitted from the source  passes through 
a Faraday isolator
 (so the filter cannot 
affect the source) and then through the filter. It is therefore 
possible, by tracing over both the bath and the filter cavity, to 
write down a master equation for the source alone.

We consider as a frequency filter a two-sided Fabry-Perot cavity that can reflect and 
also transmit light. The coupling strength $\kappa$ is assumed to be 
the same 
at each mirror. The transmitted light will be in a Lorentzian 
shaped band of width $\kappa$ about the resonance frequency of the 
cavity, $\nu$. This light is 
detected by a photodetector as is the light reflected from the other 
mirror. 

The reason that the non-Markovian quantum trajectory for the 
evolution of the atom can generate the 
correct probabilities for the spectral detection without explicitly 
modeling the filter cavity (with a Hamiltonian) is  because  it is 
possible to 
eliminate the filter cavity mode completely and write the output from 
the cavity in terms of the input \cite{dans book}.

We define two output channels, $c^{\rm out}_{R}$ and 
$c^{\rm out}_{T}$, as the two output fields from the reflecting and 
transmitting mirrors, respectively, after the output 
from the atom $b^{\rm out}$  has 
interacted with the filter cavity,
\begin{eqnarray}
	c^{\rm out}_{R}(t) & = & U^{\dagger}_{F}(t,t_{0})[b^{\rm 
out}(t)+c_{T}^{\rm in}(t)]U_{F}(t,t_{0}),\\
  & = & b^{\rm out}(t)+
	c_{T}^{\rm in}(t)+\sqrt{\kappa}a(t), \\
		c^{\rm out}_{T}(t) & = & U^{\dagger}_{F}(t,t_{0})[b^{\rm 
		out}(t)+c_{T}^{\rm in}(t)]U_{F}(t,t_{0}),\\
 & = & b^{\rm out}(t)+c_{T}^{\rm in}(t)+
	\sqrt{\kappa}a(t), \label{dd}
\end{eqnarray}
where $U_{F}$ is the unitary evolution for the interaction of the 
baths with the filter defined by
\begin{equation}
	U_{F}(t,t_{0})=T\exp\left\{i\sqrt{\kappa}
	\int^{t}_{t_{0}}ds\left(a_{I}(s)[c_{T}^{\dagger{\rm 
	in}}(s)+b^{\dagger{\rm out}}(s)]-H.c.\right)\right\},
\end{equation}
where $a_{I}(s)$ is the cavity mode in the interaction picture,  $c_{T}^{\dagger{\rm in}}(s)$ is the free transmitted field and 
$b^{\rm out}(s)$ is the output field that is coupled 
to the atom, both are delta correlated.

We can now easily solve the Heisenberg equations of motion for the 
filter cavity mode by Fourier 
transforms. If we assume the state of the filter in the distant 
future is the 
vacuum state we can put $a(t_{1})=0$,
and we find
\begin{eqnarray}
	c^{\rm out}_{R}(t) & =& 
\frac{1}{\sqrt{2\pi}}\int^{\infty}_{-\infty}d\omega
\frac{i(\omega-\nu)b(\omega,t_{1})+\kappa c_{T}(\omega,t_{1})}
{\kappa-i(\omega-\nu)}e^{-i\omega(t-t_{1})}, \\
		c^{\rm out}_{T}(t) & =& 
		\frac{1}{\sqrt{2\pi}}\int^{\infty}_{-\infty}d\omega
\frac{i(\omega-\nu)c_{T}(\omega,t_{1})+\kappa b(\omega,t_{1})
}{\kappa-i(\omega-\nu)}e^{-i\omega(t-t_{1})} , 
\end{eqnarray}
These are the channel field operators corresponding to 
photodetections of the 
reflected and transmitted fields of the filter. The presence of the 
 bath $c_{T}$ preserves commutation relations.
The impulse response for each of these channels are therefore,
\begin{eqnarray}
	h^{\rm filter}_{R}(t-s) & = & 
\frac{1}{\sqrt{2\pi}}\int^{\infty}_{-\infty}d\omega\frac{i(\omega-\nu)
	}{\kappa-i(\omega-\nu)}e^{-i\omega(t-s)} , \\
 & = & -\delta(t-s)+h^{\rm filter}_{T}(t-s),\\
 h^{\rm filter}_{T}(t-s) & = & \frac{1}{\sqrt{2\pi}}\int^{\infty}_{-\infty}d\omega\frac{\kappa}
	{\kappa-i(\omega-\nu)}e^{-i\omega(t-s)}, \\
 & = & u(t-s)\kappa e^{-i\nu (t-s)} e^{-\kappa(t-s)}, 
	\end{eqnarray} 
where $u(\tau)$ is the unit step function, zero for $\tau<0$. Note 
that in line with the general discussion in the introduction if we 
detect a photon in the channel $c^{\rm out}_{T}$ then we know its 
frequency to within the 
cavity linewidth $\kappa$ but have lost information about the time of 
emission 
to within $1/\kappa$.

We can simulate the spectral measurement process via 
Eq.(\ref{probeforphotodetection}) by simply substituting in the 
filter impulse responses. The perturbation expansion in this case is in orders of 
$\gamma/\kappa$ as $\kappa$ defines the decay time of the impulse 
response functions.

\begin{figure}
\begin{center}
\epsfig{file=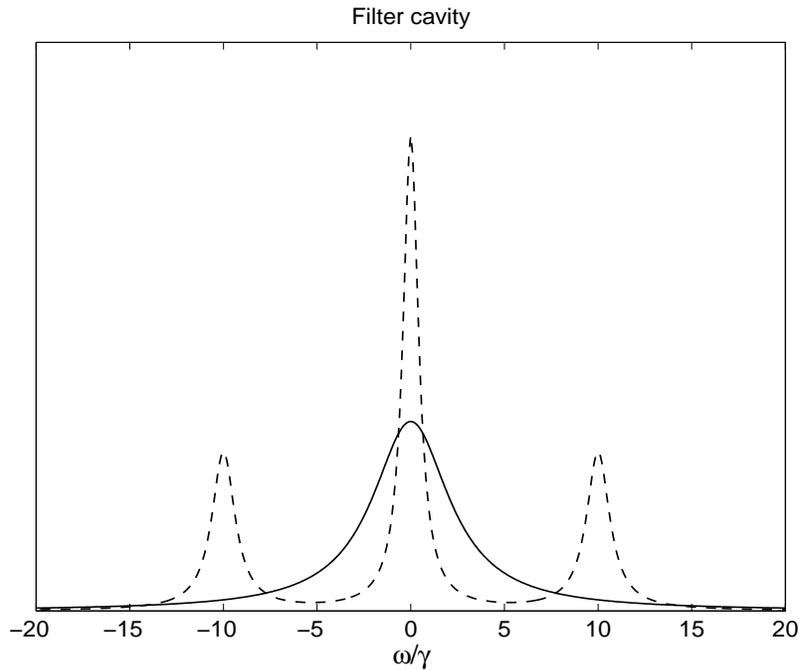,height=0.5\linewidth,width=0.6\linewidth}
\end{center}
\caption{\protect\footnotesize Schematic of	the	filtering situation	considered.	The	three peaked
spectrum of	the	two	level atom is shown	and	superimposed on	top
 is	the	Lorentzian lineshape of	the	Fabry Perot	filter.	The	filter cavity
 is	on resonance with the atomic
 transition	 $\nu=0$ (the
  central peak of the Mollow spectrum) and has a line width	larger than
   the atomic decay	rate,
  $\kappa=5\gamma$.	The	side peaks
  of the Mollow	spectrum are at	$\pm \Omega$, where	$\Omega=10\gamma$. \label{filterspectrum}}
\end{figure}

Fig.\ref{filterspectrum} is a schematic of the situation in frequency 
space, showing the 
three peaked output and a superimposed cavity filter centered on the 
central peak. In this 
situation the central peak is transmitted while the two side peaks are 
reflected by the filter. 

\begin{figure}
\begin{center}
\epsfig{file=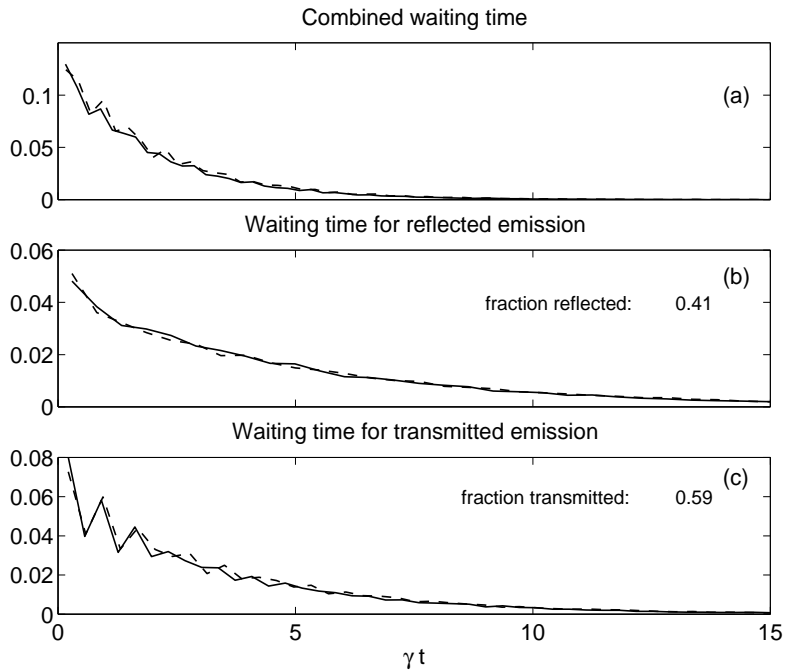,height=0.5\linewidth,width=0.6\linewidth}
\end{center}
	\caption{\protect\footnotesize Plot	of the results of a	simulation of the filtering
	situation depicted in Fig.\ref{filterspectrum},	all	three plots	are
	waiting	time distributions (in arbitrary units)
	determined by bining the times between each	subsequent detection in
	100	bins.
	The	solid line is from the non-Markovian quantum
	trajectory,	the	dashed line	is the Markovian method	for
	the	extended system.  {(a)}	is a plot
	of the combined	photodetection waiting time	distribution from both the reflected
	and	transmitted	emission. {(b)}	shows the waiting time
	distribution for just the
	reflected emission (which consists mostly of the two side peaks).
	{(c)} is a plot	of the waiting time	distribution  for the
	transmitted	emission (mostly from the central peak).The	ratio of the
	transmitted	and	reflected emission is also given.	\label{filterresults}}
\end{figure}

The results of a simulation with the parameters of 
Fig.\ref{filterspectrum} are shown in Fig.\ref{filterresults} where we 
have plotted the photodetection waiting times for the reflected and 
transmitted emission for one run of a total of $12\times 10^{4}$ detections. 
The time average of a single realization is equivalent to an ensemble 
average. The transmitted light is mostly from the central peak and the 
reflected from the side peaks. Notice that both these waiting-time 
distributions show a marked increase in the frequency of longer waiting times 
compared to the distribution for the combined emission. As 
$\gamma\sim\kappa$ the resolution of the spectral detection was of the 
order of the decay linewidth. This is reflected in the fact that a
non-vanishing fraction of the photons were detected within a memory 
time $T_{\rm m}=1/\gamma$ of each other. 

\begin{figure}
\begin{center}
\epsfig{file=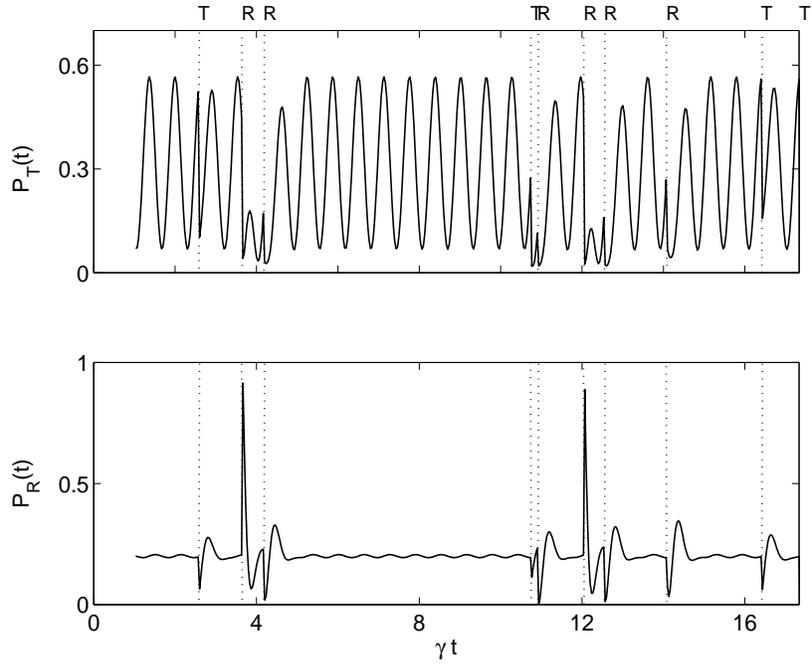,height=0.5\linewidth,width=0.6\linewidth}
\end{center}
	\caption{\protect\footnotesize Plot	of the probabilities for
	detecting a	transmitted	$P_{T}$	and	reflected $P_{R}$ photon during	the	evolution of a single quantum trajectory
	for	the	first ten detections. The labels $R	$ for reflected	and	$T$
	for	transmitted	at the top of the graph
	denote the channel that	the	detection occurred in and the dotted
	vertical lines indicate	the	time of	the	detections.\label{filtertraj}}
\end{figure}

 A plot of the evolution of the probabilities of the two channels for 
 the first ten detections of a single trajectory is shown in 
 Fig.\ref{filtertraj}. The probability of a transmission is 
 proportional to the expectation value of the number of photons in the
  filter cavity. When 
 there are no detections it 
 oscillates at the frequency $\Omega$, in time with the oscillation of 
 $\langle\sigma_{y}(t)\rangle$. The probability of detecting a reflected 
 photon is an interference 
 between the possibility of a photon  being reflected directly off the 
 mirror and the possibility of a photon coming back out of the  cavity 
 through the reflecting mirror. The oscillations are suppressed by this 
 interference. The possibility of detecting a reflected photon
  is very sensitive to the phase 
 between the states representing these two possibilities, an example is the large peaks in the reflected probability when a photon is 
 detected in the reflected channel and the oscillation has the right 
 phase.  
 
\begin{figure}
\begin{center}
\epsfig{file=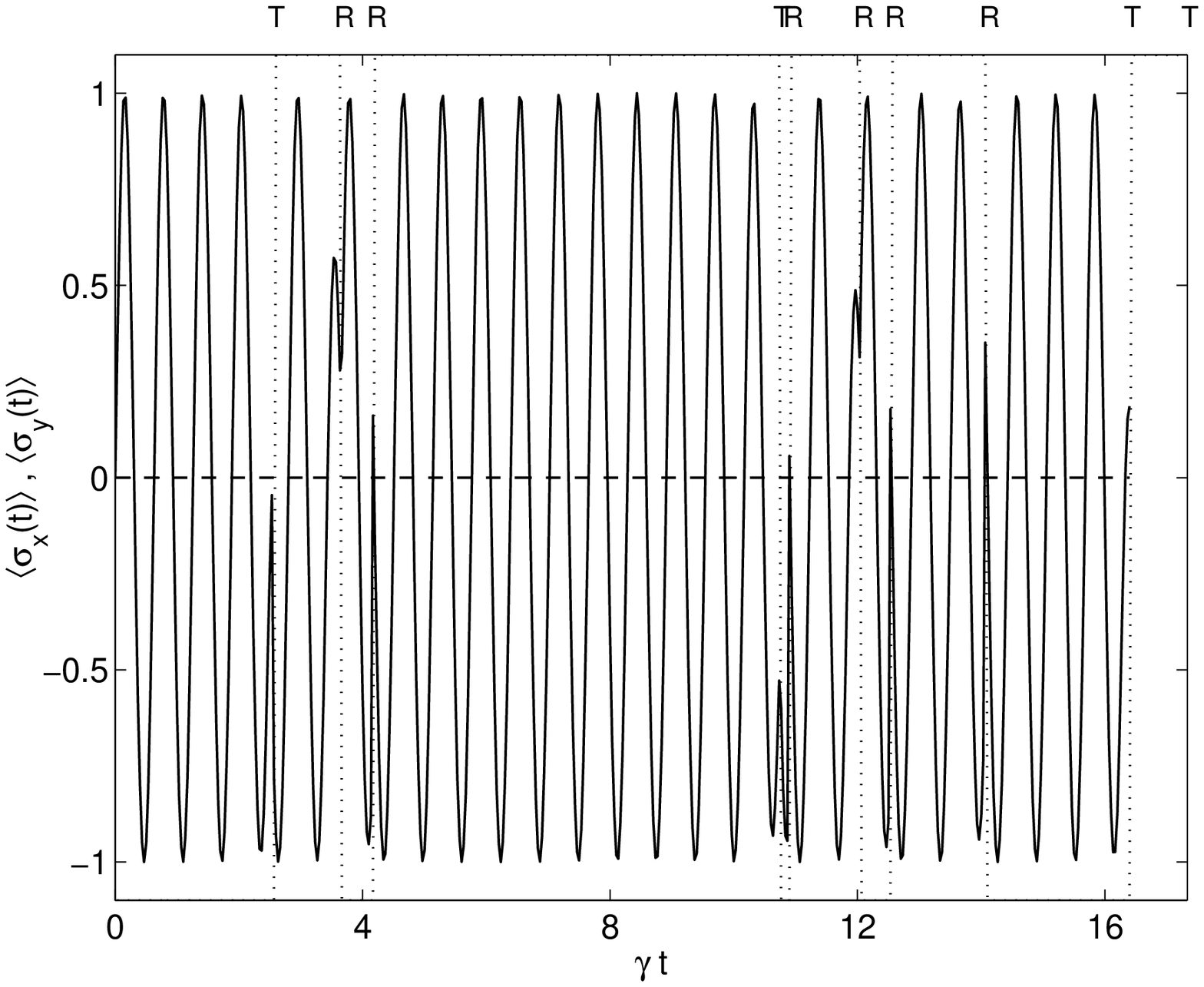,height=0.5\linewidth,width=0.6\linewidth}
\end{center}
	\caption{\protect\footnotesize
	 Plot of  the conditioned expectation values for
	$\sigma_{y}$ (solid	line) and
	$\sigma_{x}$ (dashed line) for the same	trajectory as in
	Fig.\ref{filtertraj}.\label{filtertraj2}}
\end{figure}

 Fig.\ref{filtertraj2} is a plot of the time 
 evolution of the conditioned evolution of the 
 expectation value of $\sigma_{x}$ and $\sigma_{y}$ 
 during the same trajectory.  Note that the conditioned system
 shows signs of having started to emit even before a detection occurs. 
 This is clearest in 
 the decaying $\langle\sigma_{y}(t)\rangle$ oscillation before the second and sixth 
 detection. This is a clear indication of the non-Markovian behavior 
 of the conditioned state; measurement probabilities depend on the 
 previous state of the system and the conditioned state of the system 
 is affected 
 by a detection at a later time. The interpretation of this 
 behavior goes as follows: at the time of detection the system 
 + bath is in a superposition of all the possible states, a detection 
  selects out the state that corresponds to one particle in the bath;
  associated with this state is a particular history, i.e., a weighted 
  sum of emission times; this 
 history then determines the conditioned state. 
 In the filter cavity case $\sigma_{x}$ is a constant of the free motion and is 
 unaffected by the measurement process as
 the measurement does not distinguish between the sidebands.

This situation can also be written in terms of the theory of 
cascaded systems \cite{gardinercascade,carmichaelcascade}. In that 
case, the evolution of the filter cavity as 
well as the atomic system are treated together as a coupled system. This is a special case of the  more general 
idea that was mentioned in the introduction. The atom and the filter are coupled to the reflected bath at the same 
physical 
location and they are also coupled together directly.
The quantum trajectory is then evolved by simulating the conditioned 
evolution of the complete system of source and filter.
This cascaded system trajectory can be described by the effective 
Hamiltonian (in the Schr\"{o}dinger picture),
\begin{equation}
	H_{\rm eff}=H_{\rm atom}+(\nu-i\kappa)a^{\dagger}a
	-i\left(\frac{\gamma}{2}\sigma^{\dagger}\sigma+\sqrt{\gamma\kappa}\sigma 
	a^{\dagger}\right),
\end{equation}
with the collapse operators
\begin{eqnarray}
	C_{T} & = & \sqrt{\kappa}a,\\
	C_{R} & = & \sqrt{\kappa}a+\sqrt{\gamma}\sigma.
\end{eqnarray}
Because we can formulate this situation in two different ways (in 
terms of normal quantum trajectories for an extended system, and in 
terms of a non-Markovian trajectory for the source alone), comparing 
the results of the two simulations provides a good check of the 
non-Markovian trajectory. In fact if one was interested in this 
particular system simulating the evolution of the extended system is 
much simpler and requires much less computer time than the 
non-Markovian equivalent.  

In Fig.\ref{filterresults} we have plotted the various waiting times 
taken from simulations of both the non-Markovian method and the  
 Markovian method for the extended-system the results show good agreement, with the discrepancy 
within the statistical fluctuation. In general, the non-Markovian simulation takes 
of the order of 10 times as long to run on a computer. 

In the next section we will give 
an example where it is not obvious how to build an extended system 
that would accurately simulate the measurement situation.

\subsubsection{The Prism}
\begin{figure}
\begin{center}
\epsfig{file=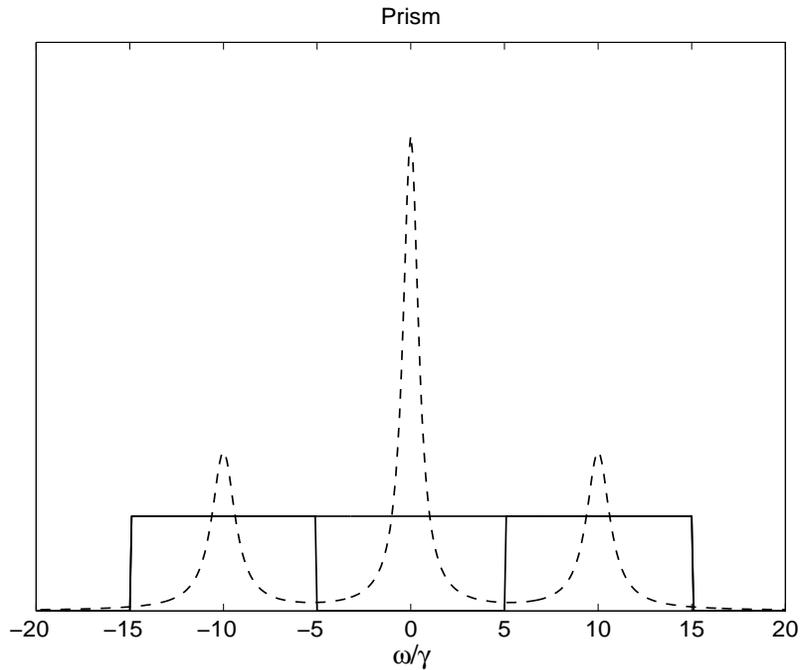,height=0.5\linewidth,width=0.6\linewidth}
\end{center}
\caption{\protect\footnotesize Schematic of	the	prism-like spectral	detection situation.
 The three peaked spectrum of the two level	atom is	shown and superimposed
 on	top	are	the	top-hat	functions corresponding	to the detectors centered on
 each peak.	 The spread	of the frequency band of
  each channel is  $\Delta/2=5\gamma$ and $\omega_{L}=-\Omega$,
  $\omega_{R}=+\Omega$ and $\omega_{C}=0$,	where, as before, $\Omega=10\gamma$.
 \label{prismspectrum}}
\end{figure}

We consider a simple model of spectral detection performed by a 
prism, where the light emitted from a source propagates through a 
prism (or a spectral grating) which spreads the light into a spectrum and is
 then incident on 
an array of photodetectors. Each detector is
then effectively sensitive to a sharply defined band of frequencies. We can then 
model this situation by assigning a top-hat 
frequency response function to each photodetector (labeled by the 
variable $n$) centered about a frequency $\omega_{n}$. 
The output channel field is given by
\begin{equation}
c^{\rm out}_{n}(t)=\frac{1}{\sqrt{2\pi}}\int^{\omega_{n}+\Delta/2}_{\omega_{n}-\Delta/2}d\omega
b(\omega,t_{1})e^{-i\omega(t-t_{1})}+\mbox{other baths},
\end{equation}
where $\Delta$ is the width of the band (which we have assumed are 
the 
same for all $n$). 
The prism then has the  impulse response functions
\begin{equation}
	h^{\rm prism}_{n}(t-s)=\frac{2e^{-i\omega_{n}(t-s)}}{\sqrt{2\pi}}
	\frac{\sin\Delta(t-s)}{t-s}.
\end{equation}
These response functions do not obey causality. This is because 
we have not included the propagation time from the system to the 
detector. A finite approximation (which is physically valid as an 
infinitely sharp cut in frequency is non-physical) to this will be nonzero
 over a time 
interval of $[T_{\rm m}/2,-T_{\rm m}/2)$. We can then take the greatest 
time of 
this interval as the time of a detection. 
The prism example has the advantage that we can easily model the 
detection of 
the three peaks of the Mollow spectrum by assigning a channel field 
to 
each of the peaks. We consider a spectrometer that splits the light 
into three frequency bands with each band centered on a separate peak of 
the Mollow spectrum.
A schematic of the situation is shown in 
Fig.\ref{prismspectrum}.

\begin{figure}
\begin{center}
\epsfig{file=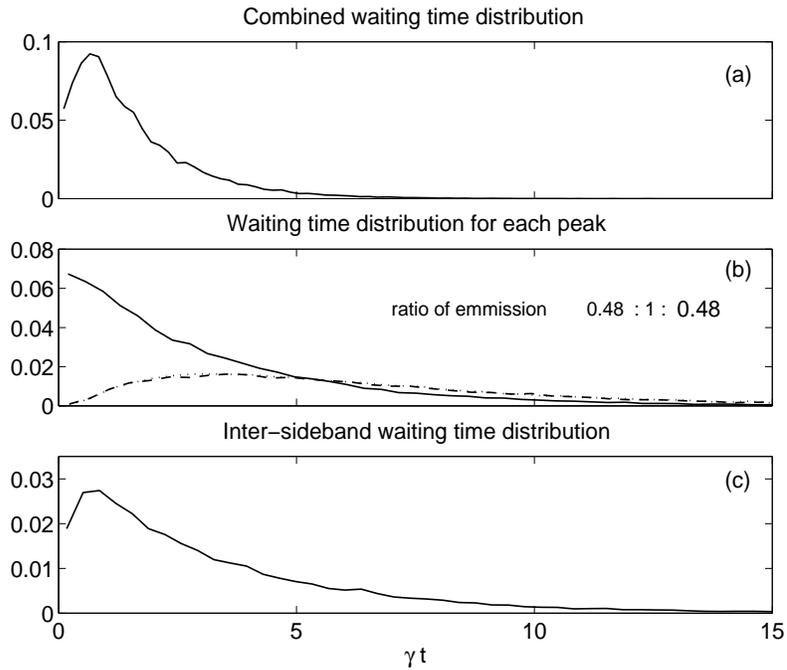,height=0.5\linewidth,width=0.6\linewidth}
\end{center}
	\caption{\protect\footnotesize Plot	of the results of a	simulation of the prism
	 depicted in Fig.\ref{prismspectrum}, all three	plots are
	waiting	time distributions	determined by bining the times between each
	subsequent detection in	100	bins.
	{(a)} is a plot	of the combined	photodetection waiting time	distribution from all
	three spectral bands, (compare with	Fig.\ref{filterresults}	{(a)}).
	{(b)} shows	the	waiting	time
	distributions within each spectral band	(each band coinciding with
	one	peak of	the	emission). The solid line corresponds to the central
	peak and the dashed	and	dotted lines to	the	side-peaks.	{(c)} is a
	 plot of the inter-sideband	waiting	time distribution, i.e., the time
	 taken between a detection in one of the side peaks	and	a detection	in
	 the other.	The	ratio of the emission into each	peak is	also shown\label{prismresults}}
\end{figure}

The results of a computer simulation of $8\times 10^{4}$ detections are shown in 
Fig.\ref{prismresults}, with system parameters as in 
Fig.\ref{prismspectrum} and with $T_{\rm m}=1/\gamma$. The results show a definite anti-bunching of 
the side-peak photons (characterized by the peak of the distribution 
being shifted to longer waiting times). They also demonstrate that the inter-sideband 
waiting times (i.e., the time between an emission into one sideband and 
an 
emission in the other) have a distribution similar to that of the central 
peak. Note that this plot shows details of the waiting-time 
distributions closer 
to  zero than the filter case.

\begin{figure}
\begin{center}
\epsfig{file=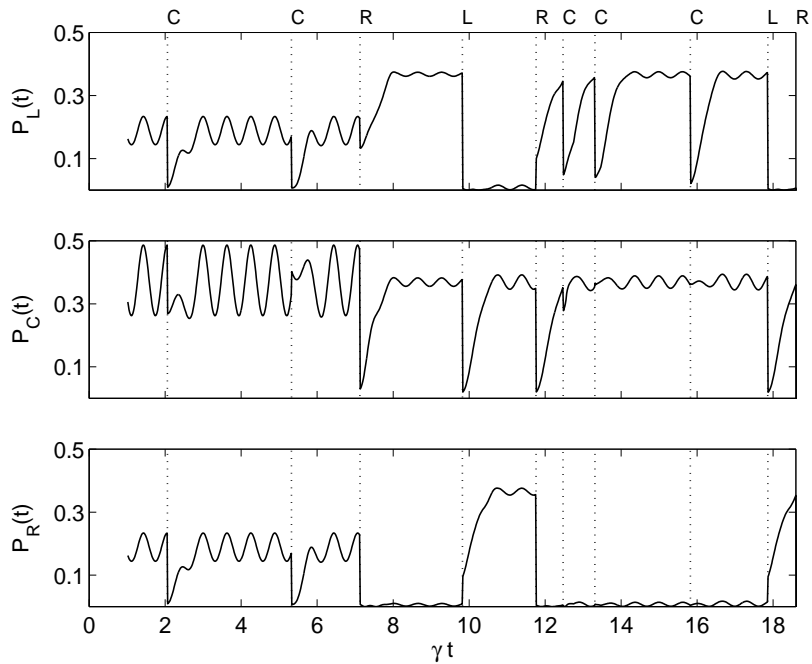,height=0.5\linewidth,width=0.6\linewidth}
\end{center}
	\caption{\protect\footnotesize Plot	of the probabilities for
	detecting a	photon from	the	central	peak $P_{C}$
	and	the	two	side peaks $P_{L}$ and $P_{R}$
	during the evolution of	a single quantum trajectory
	for	the	first ten detections. The labels $C$ for central and $L$
	and	$R$
	for	the	side-peaks at the top of the graph
	denote the channel that	the	detection occurred in and the dotted
	vertical lines indicate	the	time of	the	detections.	\label{prismtraj}  }
\end{figure}

\begin{figure}
\begin{center}
\epsfig{file=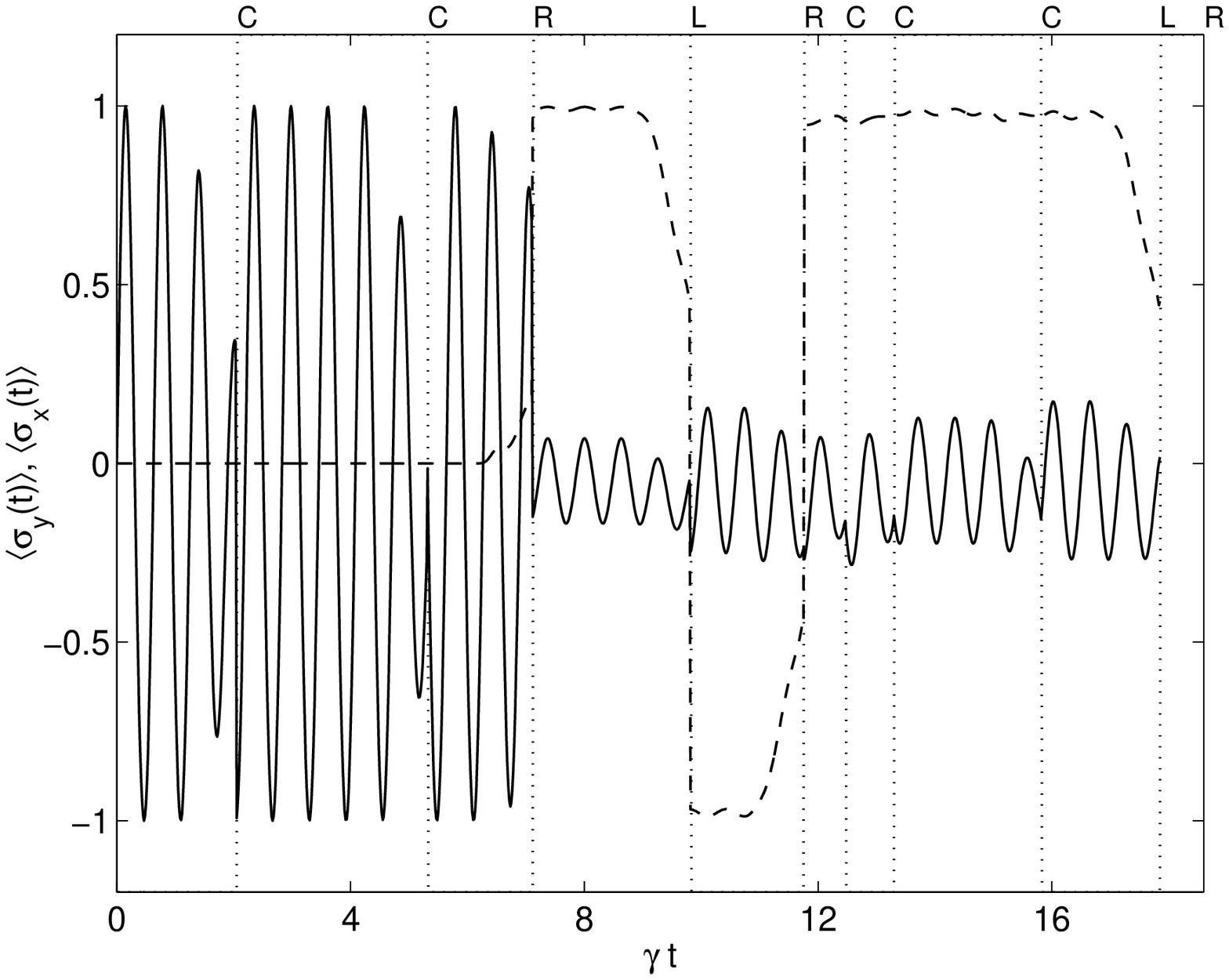,height=0.5\linewidth,width=0.6\linewidth}
\end{center}
	\caption{\protect\footnotesize
	Plot of	the	conditioned	expectation	values of
	$\sigma_{y}$ (solid	line) and
	$\sigma_{x}$ (dashed line) for the same	trajectory as depicted in
	Fig.\ref{prismtraj}. \label{prismtraj2}}
\end{figure}

A plot of the detection probabilities in each channel for one run of 
ten detections is shown in Fig.\ref{prismtraj}. We have labeled
the channels; L (left), R (right) and C (center), referring to the side and center peaks of the 
Mollow spectrum that the channels correspond to.
For our purposes the thing to note about these trajectories is that after 
a detection 
occurs in the right (left) channel the probability to get another 
detection in the same channel is nearly zero until a detection 
occurs in the left (right) channel regardless of detections that 
occur in the center channel. This is just the predicted anti-bunching 
in the side-peaks. This phenomenon is reflected in the conditioned state of the 
atom. A plot of the expectation values of  $\sigma_{x}$ and  $\sigma_{y}$ for this conditioned state 
are shown in Fig.\ref{prismtraj2}. Whereas in the cavity filter 
case $\langle\sigma_{x}(t)\rangle$ was zero 
throughout the trajectory, here, because the measurement process can 
distinguish the side-peaks, the atom gets projected into the 
eigenstates of $\sigma_{x}$ with each measurement of a side-peak 
photon. The amplitude of oscillations in $\langle\sigma_{y}(t)\rangle$ 
decrease when $|\langle\sigma_{y}(t)\rangle|$ is large as the atoms behavior is basically restricted to the surface 
of the Bloch sphere. Note that the conditioned expectation values $\langle\sigma_{y}(t)\rangle$ and
 $\langle\sigma_{x}(t)\rangle$ show even more obvious signs of future 
 detections than the filter cavity case, for example, the very obvious 
 decay of the oscillations of $\langle\sigma_{y}(t)\rangle$ and the motion of 
 $\langle\sigma_{x}(t)\rangle$ towards zero long before a detection. This is more 
 pronounced in the prism case because the impulse response is 
 a sinc function as opposed to a more rapidly decaying exponential 
 in the filter case.

\section*{Conclusion}
We have derived a general form for non-Markovian quantum 
trajectories 
corresponding to a particle conserving coupling between the system 
and bath. We have also introduced the concept of a complete set of 
measurement 
channels to model measurement devices that are sensitive to only a 
range of modes of the bath.
In the limit of weak coupling the non-Markovian quantum trajectories 
so derived are amenable to computer 
simulations. We have demonstrated the practicality of the method by 
making 
computer simulations of the special case of optical spectral 
detection. 
The results agree with the predictions of more 
traditional Markovian methods.
The theory is  general enough to be applied to a number 
of areas in physics where open systems arise and where the Markovian 
assumption cannot be made.
In further work we hope to apply the above theory to 
simulating the statistics of the output coupled atoms from a 
Bose-Einstein condensate \cite{hope,savage}.  Another topical problem to 
which these methods 
could be applied is  that of radiation 
into band-gap materials  \cite{bay,vats}. 

\vspace{1cm}				 
			
M.J. would like to thank J. Ruostekoski for helpful discussions and S. M. 
Tan and M. Steel for helpful advice about the numerical simulations. The 
authors are grateful for the support of the Marsden Fund of the Royal 
Society of New Zealand and the University of Auckland Research Fund.

\newpage

\large{FIGURES}

\begin{thebibliography}{99}
\bibitem{dalibar}J. Dalibard, Y. Castin, and K. M\o lmer, Phys. Rev. 
Lett. {\bf 68}, 580 (1992).
\bibitem{carmichael}  H. Carmichael, {\em An Open Systems Approach to Quantum Optics}, 
Lecture Notes in Physics m18 (Springer-Verlag, New York,
 1991).
\bibitem{gisin} N. Gisin and I. Percival, Phys. Lett. {\bf 167A} 
(1992).
\bibitem{gardiner}  C. W. Gardiner A. S. Parkins and P. Zoller Phys. Rev. A,
 {\bf 46}, 4363 (1992).
\bibitem{molmer} K. M\o lmer, Y. Castin and J. Dalibard, J. Opt. Soc. 
Am. B {\bf 10}, 524 (1993).
\bibitem{dum} R. Dum, P. Zoller and H. Ritsch, Phys. Rev. A {\bf 45}, 
4879 (1992).
\bibitem{tian} L. Tian and H. J. Carmichael, Phys. Rev. A, {\bf 46}, 
R6801 (1992).
\bibitem{photon} M. D. Srinivas and E. B. Davies,  Opt. Acta {\bf 28}, 981 (1981).
\bibitem{wiseman} H. M. Wiseman and G. J. Milburn, Phys. Rev. A {\bf
47}, 642 (1993).
\bibitem{goetsch} P. Goetsch and  R. Graham, Phys. Rev. A {\bf 50}, 
5242 (1994).
\bibitem{javanainen} J. Javanainen and S. M. Yoo Phys. Rev. Lett. 
{\bf 76}, 161 (1996).
\bibitem{bay}S. Bay, P. Lambropoulos and K. M\o lmer, Phys. Rev. 
Lett. {\bf 79}, 14 (1997).
\bibitem{vats} N. Vats and S. John, (unpublished).
\bibitem{outputcoupler} M. O. Mewes {\em et al}., Phys. Rev. Lett. 
{\bf 78}, 582 (1997).
\bibitem{hope} J. Hope, Phys. Rev. A {\bf 55}, R2531, (1997).
\bibitem{savage} G. M. Moy, J. J. Hope and C. M. Savage, (unpublished).

\bibitem{caves} C. M. Caves, Phys. Rev. D {\bf 35}, 1815 (1986).
\bibitem{barchielli} A. Barchielli, Phys.Rev. D {\bf 34}, 2527 (1986).

\bibitem{wisemanthesis} H. M. Wiseman, Ph.D. Thesis, University
of Queensland (1994).
\bibitem{cohen}C. Cohen-Tannoudji and S. Reynaud,
Phil. Trans. R.  Soc. Lond. A {\bf 293}, 223 (1979).
\bibitem{cresser} J. D. Cresser, J. Phys. B: At. Mol. Phys. {\bf 20}, 
4915 (1987).
\bibitem{wisemanhetrodyne}  H. M. Wiseman and G. J. Milburn, Phys. Rev. 
A {\bf 47}, 1652 (1993).
\bibitem{imamoglu} A. Imamo\={g}lu, Phys. Rev. A {\bf 50}, 3650 (1994).
\bibitem{diosi97} L. Di\`{o}si and W. T. Strunz, Phys. Lett. A {\bf
235},  569 (1997).
\bibitem{diosi98} L. Di\`{o}si, N. Gisin and W. T. Strunz, (unpublished).

\bibitem{mollow} B. R. Mollow, Phys. Rev. {\bf 188}, 1969 (1969).
\bibitem{aspect} A. Aspect, G. Roger, S. Reynaud, J. Dalibard, and C. 
Cohen-Tannoudji, Phys. Rev. Lett. {\bf 45}, 617 (1980).
\bibitem{gardinercascade} C. W. Gardiner, Phys. Rev. Lett. {\bf 70}, 
2269 (1993)
\bibitem{carmichaelcascade} H. J. Carmichael, Phys. Rev. Lett. {\bf 
70}, 2273 (1993). 
(1996).
\bibitem{davies} E. B. Davies, {\em Quantum Theory of Open Systems}, 
(Academic Press, London, 1976).
\bibitem{qn} C. W. Gardiner, {\em Quantum Noise}, Springer series in synergetics 
(Springer-Verlag, New York,1991).
\bibitem{collett} C. W. Gardiner and M. J. Collett, Phys. Rev. A {\bf 31}, 
3761 (1985).
\bibitem{digital} S. D. Stearns and D. R. Hush, {\em Digital Signal 
Analysis 2nd ed.}, Prentice Hall Signal Processing Series, (Prentice 
Hall, New Jersey, 1990)

\bibitem{dans book} D. F. Walls and G. J. Milburn, {\em Quantum
Optics}  (Springer-Verlag, New York, 1994).
\end{thebibliography}
\end{document}